\documentclass{aa}  
\usepackage[varg]{txfonts}
\usepackage[dvipsnames]{xcolor}
\usepackage{hyperref}
\usepackage{multirow}

\usepackage{verbatim}
\usepackage{float}
\usepackage{graphicx}

\usepackage{lineno}

\begin{document} 

    \title{Optical spectropolarimetry of extreme H$\alpha$ line profiles in seven active galactic nuclei}

    \titlerunning {Optical spectropolarimetry of seven extreme AGNs}
    
   \author{J. Biedermann \inst{1}\thanks{\href{mailto:julie.biedermann@astro.unistra.fr}{julie.biedermann@astro.unistra.fr}}    
            \and
            F. Marin\inst{1}
            \and 
            D. Hutsem\'ekers\inst{2}
            \and 
            C. M. Gaskell\inst{3} }

   \institute{Universit\'e de Strasbourg, CNRS, Observatoire astronomique de Strasbourg, UMR 7550, F-67000 Strasbourg, France
              \and
              Institut d’Astrophysique et de G\'eophysique, Universit\'e de Li\`ege, All\'ee du 6 Ao\^ut 19c, B5c, 4000 Li\`ege, Belgium
              \and 
              Department of Astronomy and Astrophysics, University of California at Santa Cruz, California 95064, USA }

   \date{Received month day, 2026; accepted Month Day, 2026}
   
   \abstract
   {Some active galactic nuclei (AGNs) are known to show extremely asymmetric broad Balmer lines, with their main peak redshifted or blueshifted by thousands of km~s$^{-1}$, severely challenging our understanding of the standard AGN paradigm.}
   {We wanted to explore the causes of such asymmetric features by carrying out a spectropolarimetric study of a sample of bright AGNs with known extreme Balmer line profiles.}
   {We present optical spectropolarimetry of seven bright (V $\le$ 17.5) Seyfert-1 galaxies obtained with the Very Large Telescope (VLT) FOcal Reducer/low dispersion Spectrograph 2 (FORS2) instrument from March 2012 to January 2013. The main broad Balmer lines (H$\alpha$ and $H\beta$, and H$\gamma$ occasionally) were sampled with a resolving power ranging from 350 to 500. We performed a detailed spectral decomposition of each H$\alpha$ profile to isolate kinematic substructures within the broad line region (BLR) and measured the polarization degree and angle of each component, providing constraints on the inner geometry and scattering properties.}
   {The broad H$\alpha$ components exhibit various scattering geometries. Their polarization as a function of full width at half maximum (FWHM) reveals two distinct behaviors, corresponding to either a BLR of roughly constant scale height or a stratified structure with decreasing height at higher velocities. This supports a disk-wind geometry linking equatorial and polar regions.}
   {}

    \maketitle
    \nolinenumbers

\section{Introduction}
\label{introduction}

Active galactic nuclei (AGNs) are among the most luminous and energetic sources of radiation in the Universe. Their luminosity, often greater than that of their host galaxy, comes from accretion of matter onto a supermassive black hole (SMBH), whose mass typically ranges from 10$^{6}$ to 10$^{9} M_{\odot}$ \citep{Rees_1984, Peterson_2004, Alexander_2025}. AGNs can be broadly divided into two categories \citep{Antonucci_2012}: "thermal AGNs" with Eddington ratios $\gtrsim 10^{-3}$, whose energy output is dominated by thermal emission associated with the "Big Blue Bump," and "non-thermal AGNs," with Eddington ratios $\lesssim 10^{-3}$ for which the energy output is dominated by mechanical energy in the jets. According to the most accepted model \citep{Antonucci_1993}, the matter that spirals toward the nucleus forms an equatorial accretion disk at sub-pc scales. At the outer (red) edge of the accretion flow, dense ($n \sim 10^9 - 10^{13}$ cm$^{-3}$, \citealt{Muller2020}), ionized gas clouds in Keplerian rotation form the so-called broad line region (BLR, \citealt{Gaskell_2008}). The BLR clouds are photoionized by ultraviolet (UV) continuum radiation from the central engine, producing broad emission lines (FWHM = 1000 - 25~000~km~s$^{-1}$), which are characteristic of the optical spectra of pole-on, type 1 AGNs \citep{Peterson_1993, Peterson_2004}. The BLR lies within the dust sublimation radius and is surrounded by a dusty torus that obscures the broad lines for equatorial lines of sight, leading to type-2 AGNs with only narrow emission lines from larger-scale ionized outflows.

Understanding the nature of the BLR is one of the main goals in the field of AGN research, as the BLR is an excellent probe to constrain the processes governing AGN activity. Located relatively close to the SMBH, the BLR is also a key component for estimating the mass of the central black hole. Recent studies have significantly improved our knowledge of the BLR geometry and kinematics, particularly through indirect observational techniques such as reverberation mapping (RM), which consists of measuring the time delay between variations of the UV/optical continuum and the corresponding response in the emission lines of the BLR \citep{Blandford_1982, Horne_2004} or in the re-emitted infrared (IR) continuum emission from the torus \citep{Suganuma2006}. As of today, the geometry of the BLR is generally described as a flattened distribution or disk-like structure of turbulent gas co-rotating above the accretion disk. This configuration explains both the relatively high covering factor required to reproduce the observed strength of the broad emission lines and the absence of strong Lyman continuum absorption. If the BLR had a spherical geometry, its higher covering factor would necessarily imply that the observer's line of sight would almost always intercept a significant fraction of the ionized gas. This would cause pronounced absorption of the Lyman continuum, which is inconsistent with observations. A flattened configuration thus prevails. RM studies also highlighted a radial stratification of the ionization structure \citep{Gaskell_1986, Gaskell_2011}, with highly ionized lines (such as He~II, He~I, C~IV) emitted at shorter distances from the central engine, while low ionization lines (such as Fe~II, Mg~II) originate from more distant regions. Measurement of the characteristic size of the BLR by RM studies also revealed a robust empirical relationship between its radius and the luminosity of the AGN ($R_{BLR} \propto L_{AGN}^{1/2}$) \citep{Koratkar_1991, Bentz_2006, Bentz_2009}. The outer boundary of the BLR appears to coincide with the dust sublimation radius (T$_{sub} \sim 1500$ K), implying that the BLR is the inner extension of the dusty torus \citep{Netzer_1993, Suganuma_2006, Rosborough_2025}. Spectroscopy also helped to characterize the BLR and found that, in addition to the dominant rotational movement, there is a vertical component of velocity, often interpreted as turbulence \citep{Osterbrock_1978}. A number of other spectroscopic signatures were also noticed, such as the presence of inflows and outflows, where the different components will produce different line profiles \citep{Gaskell_1988, Du_2016, Horne_2021}. Finally, by using the reverberation times and the velocity dispersion of the lines, it becomes possible to estimate the mass of the SMBH with the virial method, $M_{BH}=f_{virial}Rv^{2}/G$, where $f_{virial}$ is the virial factor, $R$ is the radius of the BLR, and $v$ represents the characteristic velocity of the BLR gas \citep{Onken_2004, Rosborough_2025}.

In order to explore the unresolved central region of AGNs, and especially the BLR, spectropolarimetry offers a complementary and particularly sensitive approach \citep{Smith_2005, Afanasiev_2014}. Because scattering-induced polarization is sensitive to the geometry of the scattering region and to the orientation of the last scattering surface, the combination of spectroscopy and polarimetry provides access to the morphology and kinematics of structures that cannot be spatially resolved with direct imaging. Polarimetric observations have already played a crucial role in the development of the unified model of AGNs \citep{Antonucci_1993}, showing that, to first order, type 1 and type 2 AGNs are intrinsically the same objects but observed at different inclinations. The discovery of polarized broad emission lines in the Seyfert~2 galaxy NGC~1068 provided direct evidence of the presence of a hidden BLR, obscured by the dusty torus and revealed only by scattered light \citep{Antonucci_1993, Antonucci_1985}. Moreover, polarimetry offers a key constrain on the scattering geometry. The orientation of the polarization position angle ($\theta$) relative to the projected angle of the radio source axis determines which scattering region dominate. When the optical polarization angle is perpendicular to the radio axis, polar scattering is dominant, as expected in type~2 AGNs. If $\theta$ is parallel to the radio axis equatorial scattering geometry dominates as observed in type~1 AGNs \citep{Smith_2004}. When used in type-1 AGNs, spectropolarimetry can detect significant variations in the polarization degree ($P$) and polarization position angle ($\theta$) across the broad emission lines -- such as H$\alpha$ --, providing insights into the geometry and velocity field of the BLR and other scattering regions \citep{Goodrich_1994, Smith_2002, Smith_2005}. In addition, polarimetry can also be used to estimate the mass of SMBHs using variations of the polarization across broad emission lines \citep{Afanasiev_2015, Savi_2018}, independently of classical spectroscopic techniques.

When viewed close to face-on, a flattened, turbulent BLR can reproduce the symmetric single-peaked “logarithmic” Balmer profiles of most UV selected AGNs (majority of AGNs in the SDSS). However, not all AGNs have simple, symmetric broad line profiles (including double-peaked Balmer lines), as expected from a purely Keplerian disk \citep{Marsh1988,Perez_1988, Eracleous_1997, Eracleous_2003}. When viewed at higher inclinations it produces broader, double-peaked profiles (see, for example, Fig~2 of \citet{Gaskell_2011}. A major limitation of this standard rotating BLR model is that it does not naturally explain the frequent and sometimes strong asymmetries observed in broad Balmer lines, including displaced peaks and red or blue wings extending over several thousand~km~s$^{-1}$. Some AGNs even display triple-peaked or more exotic Balmer profiles \citep{Benitez2019}, suggesting more complex BLR dynamics or geometries \citep{Eracleous_1994, Corbett_2000, Eracleous_2003}. Any complete BLR model must therefore accounts for these extreme profiles, and proposed explanations include binary SMBHs, off-axis disk illumination, and partial obscuration of the BLR by small dusty clouds.

\citet{Gaskell_1983} proposed that each peak corresponds to a separate BLR associated with two black holes in a supermassive binary. This model predicts that the intensities of the two peaks should vary independently and that their wavelengths should slowly change on an orbital timescale. The most prominent peak in 3C 390.3 did show changes in radial velocity consistent with orbital motion over a period of a couple of decades \citep{Gaskell_1996}, but the binary model was firmly excluded for this object and others by the simultaneous variation of the blue and red displaced peaks on a light crossing time \citep{Gaskell_1999, Gaskell_2010_2}. Despite this, the binary black hole hypothesis has remained under active consideration as an explanation for some asymmetric line BLR profiles (e.g., \citet{Eracleous_2012}) because supermassive binary black holes must form because in galaxy mergers \citep{Begelman_1980}. Moreover, it is expected that the existence of flat cores in massive elliptical galaxies are a consequence of scouring out of the inner regions of galaxies by binary black holes \citep{Ebisuzaki_19991}. Another model proposed to explain BLR peaks with large velocity shifts is that the black hole resulting from the final merger of a binary black hole is being expelled from the host galaxy by the gravitational recoil following the merger \citep{Merritt_2006, Bonning_2007}.

While the possibility of binary black holes or gravitational recoil explaining extreme BLR profiles remains interesting, it must be remembered that asymmetric BLR profiles are very common and indeed are almost universally seen among the broader profiles.  The simplest explanation of asymmetries in general is that they are due to irregularities in the BLR itself.  For example, \citet{Wanders_1996} proposed that the profile irregularities in NGC 5548 were due to the BLR being clumpy.  The problem with asymmetric BLR profiles being due to a clumpy distribution is that the BLR is rotating so Keplerian shear will soon wipe out irregularities.

\citet{Gaskell_2010, Gaskell_2011} pointed out that variability of the emission from accretion disks must necessarily be non-axisymmetric and that off-axis emission can readily explain extreme line profiles (see Fig. 3 of \citet{Gaskell_2010}).  Support for this hypothesis comes from the variability of line profiles. Flares in AGNs produce rapid variability.  Since these flares are necessarily off-axis, they produce rapid changes in just small parts of line profiles, as is commonly found (see, for example, the changes in the Balmer line profiles of NG 5548 in a very narrow velocity range over only three months shown in Fig. 1 of \citet{Denney_2009}).  If non-axisymmetric emission is producing major asymmetries that last for years or decades, the asymmetric (or anisotropic emission) must endure for years or decades.

\citet{Gaskell_2018} proposed another explanation of broad-line asymmetries: partial obscuration of the BLR by moving dust clouds.  This too can readily explain extreme line profiles (see their Fig. 2).  There has long been abundant evidence for variable partial coverage of AGNs (see references in \citet{Gaskell_2018}). Since the timescale for dust moving across the BLR is fairly long (months to years or decades), the partial obscuration hypothesis explains why large asymmetries can last for years.

To summarize, we have multiple explanations offered to explain the asymmetries in broad emission lines in AGNs. The two explanations with the strongest support (see above) are the off-axis emission hypothesis and the variable partial covering hypothesis. Both can reproduce the line profiles. The two more exotic explanations of SMBHs binary and recoiling black holes are problematic, but at present cannot be ruled out for the more extreme cases.

Distinguishing between the various proposed scenarios is complex, mainly because they involve a variety of physical mechanisms and dynamical timescales. However, all of them are based on asymmetric configuration, a perfect set-up for enhanced polarimetric signatures. The different models offered to explain strong BLR asymmetries (see above) will therefore produce different spectropolarimetric behavior.  For example, \citet{Goosmann_2014} have modeled the spectropolarimetric signatures to be expected from off-axis emission flares. However, these spectral signatures may also result from the intrinsically complex structure of the BLR, without requiring an exceptional scenario. The BLR is now widely interpreted as originating from a disk wind geometry. \citet{Elvis_2000} proposed this structure, in which a warm ($\sim$ 10$^{6}$K) and highly ionized flow is elevated from the surface of the accretion disk over a narrow range of radii and accelerated by radiation pressure. The outflow rises in a quasi vertical direction before bending outward and accelerating radially, forming a cone with an opening angle of approximately 60$^{\circ}$. Spectropolarimetric simulations have shown that this geometry successfully reproduced the polarization signatures of both type~1 and type~2 AGNs \citep{Marin_2013}. For this reason, we present spectropolarimetric observations of seven AGNs with extreme, broad, and asymmetric Balmer line profiles, focusing on H$\alpha$ due to better statistics. In Sect.~\ref{observations}, we present our sample of AGNs and detail our observational campaign. In Sect.~\ref{Analysis}, we analyze the resulting spectra and decompose the H$\alpha$ lines into individual sub-components. Sect.~\ref{discussion} provides a critical analysis of the results while Sect.~\ref{Conclusions} summarizes the main findings of our paper. 

\section{Individual case analysis and results}
\label{Analysis}
\begin{figure*}
    \centering
    \includegraphics[width=1\linewidth, trim={0cm 0.75cm 0cm 1cm}, clip]{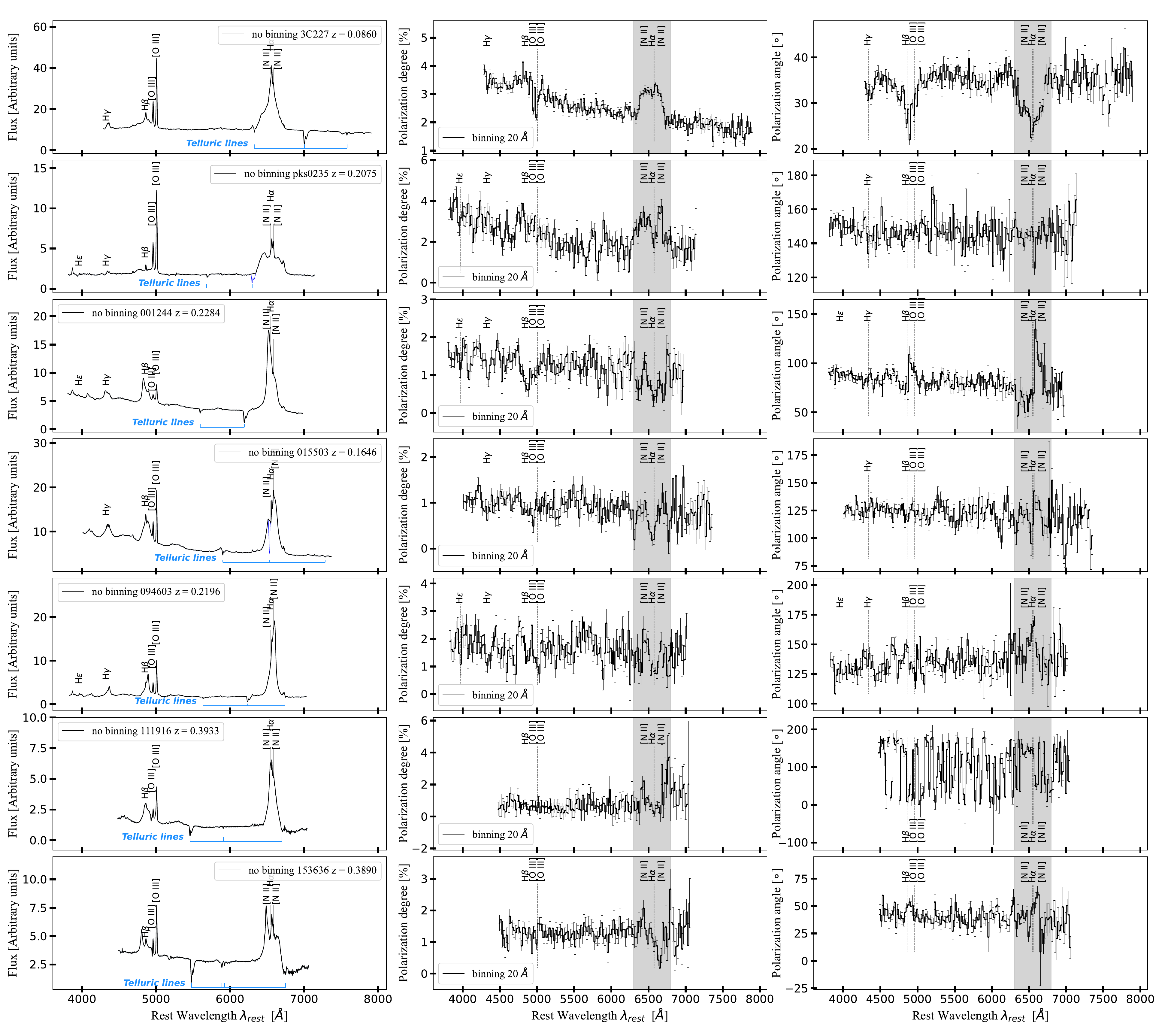}
    \caption{VLT/FORS2 spectropolarimetry of our sample of seven type 1 AGNs. First column: total flux spectrum (where tellurics are highlighted in blue), second column: polarization degree and third column: angle of polarization as a function of the rest wavelength. The gray zone, in the $P$ and $\theta$ panels, represents the waveband corresponding to the H$\alpha$ profile.}
    \label{data_all_sample}
\end{figure*}

We present the spectropolarimetric observations of 7 type~1 AGNs never published, conducted between March 2012 and January 2013 at the European Southern Observatory (ESO) using the FOcal Reducer and low dispersion Spectrograph 2 (FORS2) mounted on the UT1 of the Very Large Telescope (VLT) at Paranal Observatory. The sample of seven AGNs was initially selected through a literature survey, selecting AGNs exhibiting the most extreme velocity shift in Balmer emission lines while restricting the sample to sources that are both bright and easily observable from the southern hemisphere. Figs.~\ref{data_all_sample}, related to 3C~227, PKS 0235, SDSS 001224, SDSS 015530, SDSS 094603, SDSS 111916, and SDSS 153636, respectively, show the total flux (first column), degree of polarization $P$ (middle column) and polarization angle $\theta$ (third column), as a function of rest wavelength. The total flux spectra are shown at the native spectral resolution, while the polarized spectra are presented with a spectral binning of 20~\AA, in order to increase statistics. We show, in the appendix \ref{app:data}, the unbinned Stokes Q/I and U/I spectra plotted as a function of rest wavelength for each observation. The errors associated to each bin are indicated on the graphs.  For each object, the redshift was recalculated based on the [O~III]$\lambda\lambda$4959,5007 emission lines and are listed in Tab.~\ref{tableaux_observation}. In order to characterize the origin of the atypical line profiles, we performed a detailed spectral decomposition of each AGN spectrum in the vicinity of the H$\alpha$ emission line and detail the procedure in appendix. 

\subsection{3C~227}
\label{Analysis:3C227}

The total flux spectrum of 3C~227 is shown in Fig.~\ref{data_all_sample} (first line, first column). The spectrum is relatively flat, indicating little obscuration along the line of sight, consistent with its type-1 classification. Atmospheric absorption lines (tellurics, shown in cyan) are present but unrelated to the AGN. The very broad H$\alpha$ profile between 6300~\AA \ and 6800~\AA \ shows a slight asymmetry: the blue wing extends further and features a secondary peak around 6450~\AA, shifted by -5000~km~s$^{-1}$ relative to the narrow H$\alpha$ line, while the red wing declines smoothly. Below 5000~\AA, broad H$\beta$ (alongside [O~III] forbidden lines) and H$\gamma$ lines are seen. The polarization degree $P$ (Fig.~\ref{data_all_sample}, first line, middle column) is wavelength-dependent, ranging from $\sim$ 4\% in the blue to $\sim$ 2\% in the red, with a 1–1.5\% increase at the broad Balmer lines. Slight dips in $P$ coincide with unpolarized narrow emission lines. The polarization angle $\theta$ (Fig.~\ref{data_all_sample}, first line, third column) is ~35$^\circ$ in the continuum, rotating up to 15$^\circ$ across H$\alpha$ and H$\beta$. Similar variations in $P$ and $\theta$ suggest a common physical origin for the polarization in the Balmer-emitting regions.

\begin{figure}[!t]
    \centering
    \includegraphics[width=0.90\linewidth, trim={2cm 11cm 6.4cm 15cm}, clip]{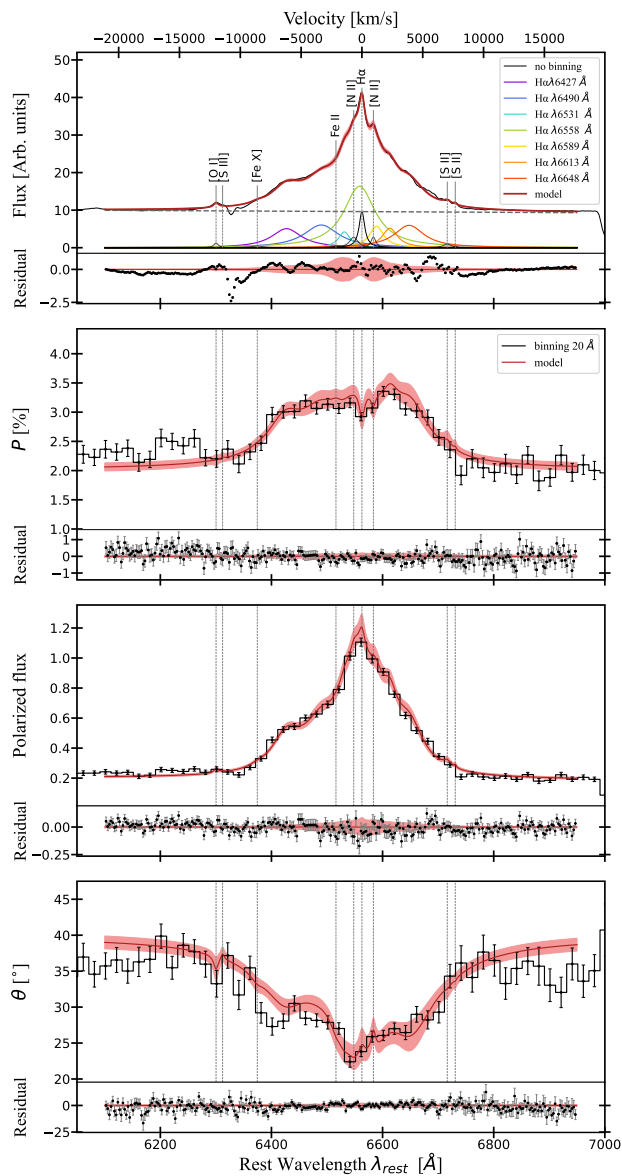}
    \caption{Spectropolarimetry decomposition of the asymmetric H$\alpha$ profile of 3~C227. From the top to the bottom, we find the total flux, the degree of polarization ($P$) and the polarized flux (that is the multiplication of the total flux with $P$) and the angle of polarization as a function of the rest wavelength in angstrom. The polarized fluxes are shown after binning of 20 \AA, while the total flux is displayed without binning. The model fitted to the unbinned data is superimposed on the observations. The residuals, shown under each of the four panels, are calculated relative to the unbinned data on which the fit was performed.}
    \label{Fig3C227_spectraldecomposition}
\end{figure}

The broad H$\alpha$ line was fitted with seven components plus the narrow lines to reproduce its ~16 000 km~s$^{-1}$ width and asymmetry (Fig.~\ref{Fig3C227_spectraldecomposition}, left). Components span a range of velocity shifts and FWHM values, reflecting complex BLR kinematics. The central component at 6558~\AA \ (blueshift 199 km~s$^{-1}$, FWHM 3130 km~s$^{-1}$) is flanked by three blueshifted components (6427, 6490, 6531~\AA) and three redshifted components (6589, 6613, 6648~\AA) with FWHM 1192–3528 km~s$^{-1}$. Polarization analysis of these components further constrains their nature. Blueshifted components at 6427 and 6490~\AA \ share similar $P$ (2.7–3.0\%) and $\theta$ (14–24$^\circ$), suggesting a common scattering geometry, while the narrower 6531~\AA \ component has lower $P$ (2.4\%) and $\theta$ = 174$^\circ$, indicating distinct scattering. Among redshifted components, 6648~\AA \ shows the highest $P$ (~3.8\%) and a $\theta$ consistent with the central component, supporting similar BLR scattering conditions.

In the protractor plot (Fig.~\ref{fig:p_theta_1}), an import diagram common to all seven AGNs from this sample, all broad H$\alpha$ components of 3C~227 are orthogonal to the AGN radio axis (~86$^\circ$, \citet{Prieto_1993_2}), suggesting a strong polar component, likely a disk wind. The observed $P$ values (2–4\%) are too high for a simple equatorial disk viewed at type-1 inclinations \citep{Goosmann2007,Marin2012}.

\subsection{PKS 0235+023}
\label{Analysis:PKS0235}

The total flux spectrum of PKS~0235 is shown in Fig.~\ref{data_all_sample} (second line, first column). The continuum is flat, indicating little dust absorption. Broad Balmer lines are present, down to H$\epsilon$, with narrow peaks superimposed. H$\alpha$ is asymmetric: the blue wing is stronger and blueshifted by $\sim$ -5100 km~s$^{-1}$ relative to the narrow line, while the red wing is redshifted by 4440 km~s$^{-1}$. Telluric absorption around 6300\AA \ prevents identification of [O~I]$\lambda$6300, [S~III]$\lambda$6312, and Si~II$\lambda$6347. The red wing shows a flux drop and rise corresponding to [S~II]$\lambda\lambda$6716,6731, while the center includes narrow H$\alpha$, [N~II]$\lambda$6548,6583, and Fe~II$\lambda$6516. The polarization degree $P$ (Fig.~\ref{data_all_sample}, second line, second column) is wavelength-dependent, from $\sim$ 4\% in the blue to $\sim$ 1\% in the red, with local variations at broad H$\alpha$. Dips coincide with the narrow line, which dilutes the polarization. Similar but weaker variations appear across H$\beta$. The polarization angle $\theta$ (Fig.~\ref{data_all_sample}, second line, third column) is mostly constant around 140$^\circ$, with a local decrease to 130$^\circ$ at H$\alpha$, interpreted as depolarization due to the narrow component.

H$\alpha$ was decomposed into six broad Lorentzian components (Fig.~\ref{FigPKS0235_spectraldecomposition}, right) to reproduce the observed asymmetry. The central component (6563\AA) is slightly blueshifted (FWHM = 1921 km~s$^{-1}$). Three blueshifted components (6425, 6468, 6517 \AA) have FWHM 2414–2946 km~s$^{-1}$, and two redshifted components (6616, 6659\AA) have FWHM 2542–3116 km~s$^{-1}$, reflecting complex BLR kinematics. Polarization analysis shows all broad components are more polarized ($P \sim$ 1.7–8.6\%) than the continuum ($P \sim$ 1.1\%). Blueshifted 6425 and 6517\AA \ components share $P$ (5.3–6.5\%) and angles (146–148$^\circ$), suggesting a common scattering geometry. The redshifted 6659\AA \ component, with the highest FWHM, has $P$ = 8.6\% and $\theta$ = 146$^\circ$, consistent with the same geometry on the opposite side. The central 6563\AA \ line has a polarization degree of about 5\% and an angle of 147$^{\circ}$, suggesting a similar scattering geometry origin to the lines mentioned previously, while the redshifted 6616\AA \ and blueshifted 6468\AA \ components display distinct angles (70$^\circ$ and 102$^\circ$) and lower $P$, possibly tracing a transition between equatorial and polar scattering regions. 

In Fig.~\ref{fig:p_theta_1} (B panel), four components have $\theta$ perpendicular to the radio axis (~56$^\circ$, \citet{Downes_1986}), consistent with polar scattering. The redshifted 6616\AA \ line aligns with the radio axis, suggesting equatorial reprocessing, while the blueshifted 6468\AA \ component has an intermediate angle, possibly probing the base of a disk-born wind with complex dynamics.

\subsection{SDSS 001224-102226}
\label{Analysis:001244}

The total flux of SDSS~001224-102226 is shown in Fig.~\ref{data_all_sample} (third line, first column). The continuum is relatively blue, indicating little reddening. Several Balmer lines are detected with broad profiles, down to H$\epsilon$. H$\beta$ and especially H$\alpha$ are very intense and display strong asymmetries, with blue wings more pronounced than the red ones. For H$\alpha$, the blue wing peaks near $\sim 6520$~\AA, while the weaker red wing lies around $\sim 6600$~\AA. A Fe~II$\lambda6516$ line is detected near the maximum intensity of the blue wing, shifted by 1900~km~s$^{-1}$ relative to the narrow H$\alpha$, possibly enhancing the asymmetry. A telluric absorption feature is present on the blue edge of the profile but does not significantly affect its morphology. The polarization spectra are shown in the central and last panels in third line of Fig.~\ref{data_all_sample}. The polarization degree is relatively constant across the spectrum, decreasing slightly from $\sim$ 1.5\% in the blue to $\sim$ 1\% in the red, possibly due to wavelength-dependent dilution by host starlight. Such nearly grey behavior is consistent with electron scattering. Local variations are nevertheless observed across the broad Balmer lines, particularly H$\alpha$ and H$\beta$. In H$\alpha$, the polarization degree shows a complex structure with three minima and two maxima. The second minimum coincides with the narrow H$\alpha$ component, while the third corresponds to the [S~II] doublet, consistent with dilution by weakly polarized narrow emission. The polarization angle remains near $\sim80^\circ$ across the continuum but varies strongly within the broad lines. In the H$\alpha$ region, $\theta$ decreases from $\sim80^\circ$ to $\sim60^\circ$, then increases sharply to $\sim140^\circ$ near the narrow component before returning to the continuum value, suggesting multiple scattering regions or complex BLR dynamics.

The H$\alpha$ spectral decomposition is presented in Fig.~\ref{Fig001244_spectraldecomposition}. The profile extends from about 6410 to 6700~\AA\ ($\sim13,000$~km~s$^{-1}$). Its asymmetry can be reproduced with three broad Lorentzian components. Two blueshifted components at 6514 and 6529~\AA\ (FWHM = 1811 and 2296~km~s$^{-1}$) dominate the blue wing. The 6514~\AA\ component lies close to Fe~II$\lambda6516$, suggesting possible contamination. The red wing is dominated by a single component at 6613~\AA\ with the largest FWHM (2689~km~s$^{-1}$), possibly originating closer to the accretion disk. The polarization properties of these components show clear contrasts. The blueshifted 6529~\AA\ component has a very low polarization ($\sim$ 0.4\%), consistent with zero within uncertainties and lower than the continuum (1.1\%), suggesting mostly direct emission. The redshifted 6613~\AA\ component shows the highest polarization ($\sim$ 2.2\%). The polarization angles also differ strongly: the blueshifted 6514~\AA\ component has $\theta \sim20^\circ$, while the 6529~\AA\ component has $\theta \sim103^\circ$, nearly perpendicular to each other, implying distinct scattering regions. The redshifted component at 6613~\AA\ has $\theta \sim151^\circ$, reinforcing its different origin.

The protractor plot for SDSS~001224 is shown in Fig.~\ref{fig:p_theta_1} (panel C). No radio axis measurement is available in the literature. The diagram nevertheless shows a gradual rotation of the polarization angle among the three broad H$\alpha$ components. In particular, the 6514 and 6529~\AA\ components are nearly perpendicular, suggesting distinct scattering geometries. These lines may trace different regions of a disk-born wind, from equatorial zones near the disk to the more extended polar BLR.

\subsection{SDSS 015530-085704}
\label{Analysis:015503}

The total flux spectrum of this source is shown in the first panel in the fourth line of Fig.~\ref{data_all_sample}. The continuum strongly decreases with wavelength, dropping by a factor of two between 4000 and 7500~\AA, which is consistent with a quasar that is not significantly reddened. Three Balmer lines are identified: H$\gamma$, H$\beta$, and H$\alpha$, all displaying broad profiles with decreasing intensity toward the blue part of the spectrum. The H$\gamma$ line exhibits a double-peaked structure with similar blue and red intensities. H$\beta$ also shows a double-peaked profile with a slightly stronger blue peak, although its red edge is contaminated by the [O~III] doublet. The H$\alpha$ profile is double-peaked as well, but with a stronger red peak. An absorption feature, likely of telluric origin, affects the blue side of H$\alpha$, impacting the Fe~II$\lambda6516$ and [N~II]$\lambda6548$ lines. The polarization degree spectrum is shown in the central panel of the fourth line in Fig.~\ref{data_all_sample}. The continuum polarization is nearly constant across the spectrum, with an average value of $\sim$ 1\%, consistent with electron scattering in a type~1 AGN. Local decreases in polarization are nevertheless observed at wavelengths corresponding to the Balmer lines, particularly H$\alpha$, where $P$ drops from $\sim$ 1\% to below 0.5\% before returning to the continuum value. The polarization angle, displayed in the lower panel, remains roughly constant around $120^\circ$, except for localized variations across the Balmer lines. In the H$\alpha$ region, $\theta$ decreases to $\sim100^\circ$ at the blue edge of the profile, then increases to $\sim140^\circ$ before returning to the continuum value, corresponding to a rotation of about $40^\circ$.

The spectral decomposition of the H$\alpha$ profile is presented in Fig.~\ref{Fig015503_spectraldecomposition}. Four broad Lorentzian components are required to reproduce the observed structure. Their FWHM values range from 1841 to 3176~km~s$^{-1}$, indicating a kinematically extended BLR. The blueshifted component at 6511~\AA\ reproduces the blue wing and has the largest width (FWHM = 3176~km~s$^{-1}$), suggesting an origin in the inner BLR. Its position also coincides with Fe~II$\lambda6516$, which may contribute to the emission. A central component at 6559~\AA\ (FWHM = 1873~km~s$^{-1}$) likely represents the main BLR emission. Two redshifted components are located at 6598~\AA\ (FWHM = 1841~km~s$^{-1}$) and 6626~\AA\ (FWHM = 2778~km~s$^{-1}$). The polarization decomposition shows distinct behaviors among these components. The blueshifted component (6511~\AA) has a very low polarization ($\sim$ 0.1\%), consistent with zero considering the uncertainties, similarly to the 6626~\AA\ component. Both are less polarized than the continuum ($\sim$ 0.9\%), suggesting predominantly direct emission. In contrast, the 6598~\AA\ component has $P \sim$ 1.9\%, and the central component (6559~\AA) shows the highest value with $P \sim$ 2.1\%. The two weakly polarized components have similar polarization angles (176$^\circ$ and 169$^\circ$), whereas the other two components show distinct orientations: 29$^\circ$ for the central component and $\sim1^\circ$ for the 6598~\AA\ component (compatible with 180$^\circ$ within uncertainties).

The polarization properties of these components are displayed in panel D of the protractor plot in Fig.~\ref{fig:p_theta_1}. No radio axis measurement is available for SDSS~015503, but the position angle of the extended NLR is $\sim5^\circ$ \citep{Husemann_2013}. Using this value as a proxy for the symmetry axis of the AGN, the central (6559~\AA) and redshifted (6598~\AA) components have polarization angles roughly parallel to the NLR axis, consistent with equatorial scattering. The blueshifted component (6511~\AA) is also compatible with this geometry within uncertainties. The remaining component, with $\theta = 29^\circ$ and $P \sim$ 2.1\%, does not align with the others and may correspond to a BLR cloud elevated above the equatorial plane, possibly lifted by radiation pressure.

\subsection{SDSS 094603+013923}
\label{Analysis:094603}

The total flux of SDSS~094603 is displayed in the first panel and the fourth in Fig.~\ref{data_all_sample}. The spectrum appears relatively flat, suggesting moderate dust absorption. Several Balmer lines are identified: H$\epsilon$, H$\gamma$, H$\beta$, and H$\alpha$. All exhibit broad profiles with a double-peaked structure. The H$\gamma$, H$\beta$, and H$\alpha$ lines show similar morphology, with a red peak more intense than the blue one, while H$\epsilon$ displays comparable intensities between the two components. The H$\alpha$ profile is the most intense Balmer line. Its red wing is dominated by a strong peak centered at 6600 \AA, with an abrupt decline on the red edge, making it clearly distinct from the [S~II] doublet at longer wavelengths. The blue side is less pronounced, although a weak structure appears around 6520 \AA. A Fe~II$\lambda$6516 component may also contribute to the asymmetry of the blue wing. The polarization degree spectrum is shown in the central panel and fourth line of Fig.~\ref{data_all_sample}. The polarization is noisy at 20~\AA\, binning, making it difficult to identify a clear continuum trend. With larger spectral bins, $P$ varies between $\sim 1.5\%$ and $\sim 2\%$, with a possible decrease toward red wavelengths, likely due to dilution by unpolarized host galaxy starlight. Despite the noise, local variations in polarization are observed at the wavelengths of the Balmer lines. In the H$\alpha$ profile, $P$ decreases from $\sim 2.5\%$ to below 1\% near the narrow component before increasing again toward the red side, consistent with dilution by narrow, unpolarized emission. The polarization angle (right panel and fourth line of Fig.~\ref{data_all_sample}) is relatively constant across the spectrum at $\sim130^{\circ}$. Local variations are nevertheless observed across the Balmer lines. In the H$\alpha$ profile, the angle increases from $\sim130^{\circ}$ to $\sim150^{\circ}$ on the blue side, reaches $\sim170^{\circ}$ near the narrow component, then decreases below $\sim120^{\circ}$ toward the red wing before returning to the continuum value. A similar but weaker pattern is seen in H$\beta$, and to a lesser extent in H$\gamma$.

Fig.~\ref{Fig094603_spectraldecomposition} (top panel) shows the spectral decomposition of the H$\alpha$ profile. The line extends over $\sim$ 7300 km~s$^{-1}$. Two broad components reproduce the observed asymmetry: a central component at 6561 \AA\ with FWHM 2327 km~s$^{-1}$, and a redshifted Lorentzian at 6603 \AA\ with FWHM 1418 km~s$^{-1}$. A Fe~II emission centered at 6516 \AA\ (FWHM = 1874 km~s$^{-1}$) likely contributes to the blue wing, although an additional H$\alpha$ component cannot be excluded. The polarimetric analysis of the broad H$\alpha$ profile (Fig.~\ref{Fig094603_spectraldecomposition}) shows similar polarization degrees for the two components, with $P\approx$ 1.7\% (6561 \AA) and $P\approx$ 1.6\% (6603 \AA). These values are lower than the continuum polarization ($\sim$ 2.3\%), suggesting dilution in the broad components. Their polarization angles are also similar, $\theta\approx$ 36$^{\circ}$ and $\theta\approx$ 46$^{\circ}$, respectively, and nearly perpendicular to the continuum angle ($\theta\approx$ 141$^{\circ}$), indicating distinct scattering regions.

The protractor plot for SDSS~094603 is shown in panel E of Fig.~\ref{fig:p_theta_1}. No radio axis information is available for this source. Nevertheless, the two broad components show similar polarization degree and angle, suggesting a common scattering geometry. Combined with the continuum polarization angle, this points to a BLR region extended toward the pole, producing orthogonal polarization relative to the continuum.

\subsection{SDSS 111916+110107}
\label{Analysis:111916}

The total flux spectrum of SDSS~111916 is shown in the first panel in the fifth line of Fig.~\ref{data_all_sample}. The continuum is relatively flat on the red side, while a slight rise toward shorter wavelengths is observed, consistent with a weakly reddened quasar. The H$\beta$ and H$\alpha$ Balmer lines are clearly identified. Both are broad and exhibit a similar asymmetry, characterized by a more intense blue wing. The H$\alpha$ profile is significantly stronger than H$\beta$. Its blue wing, centered near 6500 \AA, shows a sharp slope, while the red wing is broader and centered around 6600 \AA. The polarization spectrum is presented in the central panel in the fifth line of Fig.~\ref{data_all_sample}. The continuum polarization is roughly constant at $\sim$ 0.5\%, consistent with electron scattering. Clear polarized signatures are nevertheless visible at the wavelengths of the Balmer lines, particularly in the H$\alpha$ profile. The degree of polarization increases to about $2\%$ at the blue edge of the profile, then drops below $\sim$ 0.5\% near the center, where the narrow, unpolarized H$\alpha$ component is located. $P$ then rises again toward the red edge of the profile, reaching values above $\sim2\%$, although this region is affected by noise. The polarization angle (right panel in fifth line of Fig.~\ref{data_all_sample}) is noisy and oscillates around $0^{\circ}$ and $180^{\circ}$. For clarity, angles below $90^{\circ}$ were shifted by $180^{\circ}$. The continuum polarization angle remains relatively constant at about $150^{\circ}$. Significant variations are nevertheless observed across the H$\alpha$ profile: the angle decreases from $\sim200^{\circ}$ to $\sim150^{\circ}$ at the blue edge, remains roughly constant across the broad component, then increases to $\sim250^{\circ}$ at the narrow H$\alpha$ component before slightly decreasing toward $\sim200^{\circ}$ at the red edge.

The spectral decomposition of the H$\alpha$ profile (Fig.~\ref{Fig111916_spectraldecomposition}) requires three broad components. A blueshifted component centered at 6521 \AA\ has the largest FWHM (3148 km~s$^{-1}$), while a central component at 6539 \AA\ has the smallest FWHM (1086 km~s$^{-1}$). A redshifted component centered at 6610 \AA\ shows an intermediate FWHM of 2771 km~s$^{-1}$. Polarimetric analysis reveals different behaviors among the components. The blueshifted component (6521 \AA) has $P\sim$ 1\%, slightly above the continuum. The central component (6539 \AA) shows the highest polarization, reaching 2.7\%, while the redshifted component (6610 \AA) has an intermediate value ($P\approx$ 1.6\%). Their polarization angles also differ significantly: $128^{\circ}$ for the blueshifted component (close to the continuum value of $109^{\circ}$), $\sim178^{\circ}$ for the central component, and $56^{\circ}$ for the redshifted one, indicating that the broad H$\alpha$ components do not share a single scattering geometry.

The protractor diagram for SDSS~111916 is shown in panel F of Fig.~\ref{fig:p_theta_1}. No information on the radio axis of this source is available in the literature. The diagram nevertheless shows that the central H$\alpha$ component (6539\AA) is nearly perpendicular to the redshifted one. Overall, the different kinematic H$\alpha$ components appear to be scattered by distinct regions, suggesting a complex inner structure for this quasar.

\subsection{SDSS 153616+044127}

\label{Analysis:153636}

The total flux spectrum of SDSS~153636 is presented in the first panel in the last line of Fig.~\ref{data_all_sample}. As with SDSS~111916, the continuum appears generally flat, with a moderate blueing, suggesting weak or negligible dust absorption. Two Balmer lines are clearly identified: H$\beta$ and H$\alpha$, whose profiles have very similar morphologies. They are highly asymmetric, with a blue wing significantly more intense than the red wing or the central narrow emission. The H$\alpha$ profile shows a complex structure with three maxima: an intense blue emission centered around 6500 \AA, a weaker red peak near 6650 \AA, and a central component dominated by the narrow H$\alpha$ line and the [N~II] doublet. The noise level on the red side prevents clear identification of the [S~II] doublet. The degree of polarization is shown in the middle panel if the last line of Fig.~\ref{data_all_sample}. The continuum polarization is generally constant across the spectrum at about 1.3\%. A clear polarization structure is nevertheless observed across the H$\alpha$ profile. The degree of polarization increases to about 2\% at the blue edge, then decreases to $\sim$ 1\% across the blue wing, before rising again to $\sim$ 1.5\%. It then drops sharply to values below 0.5\% across the narrow H$\alpha$ and the [N~II] doublet, and finally returns to the continuum value on the red edge. Similar but weaker variations are visible in the H$\beta$ profile. The polarization angle (right panel, last line of Fig.~\ref{data_all_sample}) is roughly constant around 40-50$^{\circ}$ in the continuum but shows clear variations across the H$\alpha$ profile. The angle decreases to $\sim20^{\circ}$ across the blue wing, increases to $\sim60^{\circ}$ across the central component, then decreases again to $\sim20^{\circ}$ on the red edge before returning to the continuum value. A similar behavior is observed in the H$\beta$ profile with smaller amplitude.

The spectral decomposition of the H$\alpha$ profile (Fig.~\ref{Fig153636_spectraldecomposition}) requires four broad Lorentzian components. Two blueshifted components are centered at 6475 \AA\ and 6487 \AA\ with very different widths: the H$\alpha\lambda$6487 component (FWHM 1261 km~s$^{-1}$) contributes to the main blue peak, while the H$\alpha\lambda$6475 component is the widest of the profile (FWHM 4872 km~s$^{-1}$). A Fe~II line centered at 6364 \AA\ (FWHM 1177 km~s$^{-1}$) also contributes to the asymmetry of the blue wing. Two redshifted components centered at 6618 \AA\ and 6658 \AA\ reproduce the red wing, with similar widths (FWHM 2123 and 1946 km~s$^{-1}$). The polarization analysis shows that the blueshifted components exhibit low polarization degrees. The widest component (6475 \AA) has $P\approx$ 0.4\%, consistent with zero within uncertainties, while the 6487 \AA\ component has $P\approx$ 1.3\%, similar to the continuum (1.4\%). In contrast, the two redshifted components have significantly higher polarization degrees of about $P\approx$ 3\%. Their polarization angles show a wide diversity: the blueshifted components have very different angles ($100^{\circ}$ and $10^{\circ}$), while the redshifted components have angles of $99^{\circ}$ and $136^{\circ}$, suggesting a common scattering origin for the latter. However, the 6658 \AA\ component is almost perpendicular to the continuum polarization ($\theta\approx38^{\circ}$), indicating a distinct scattering region.

\begin{figure*}[!t]
    \sidecaption
    \centering
    \includegraphics[width=12cm,  trim={0cm 4.3cm 0.8cm 0cm}]{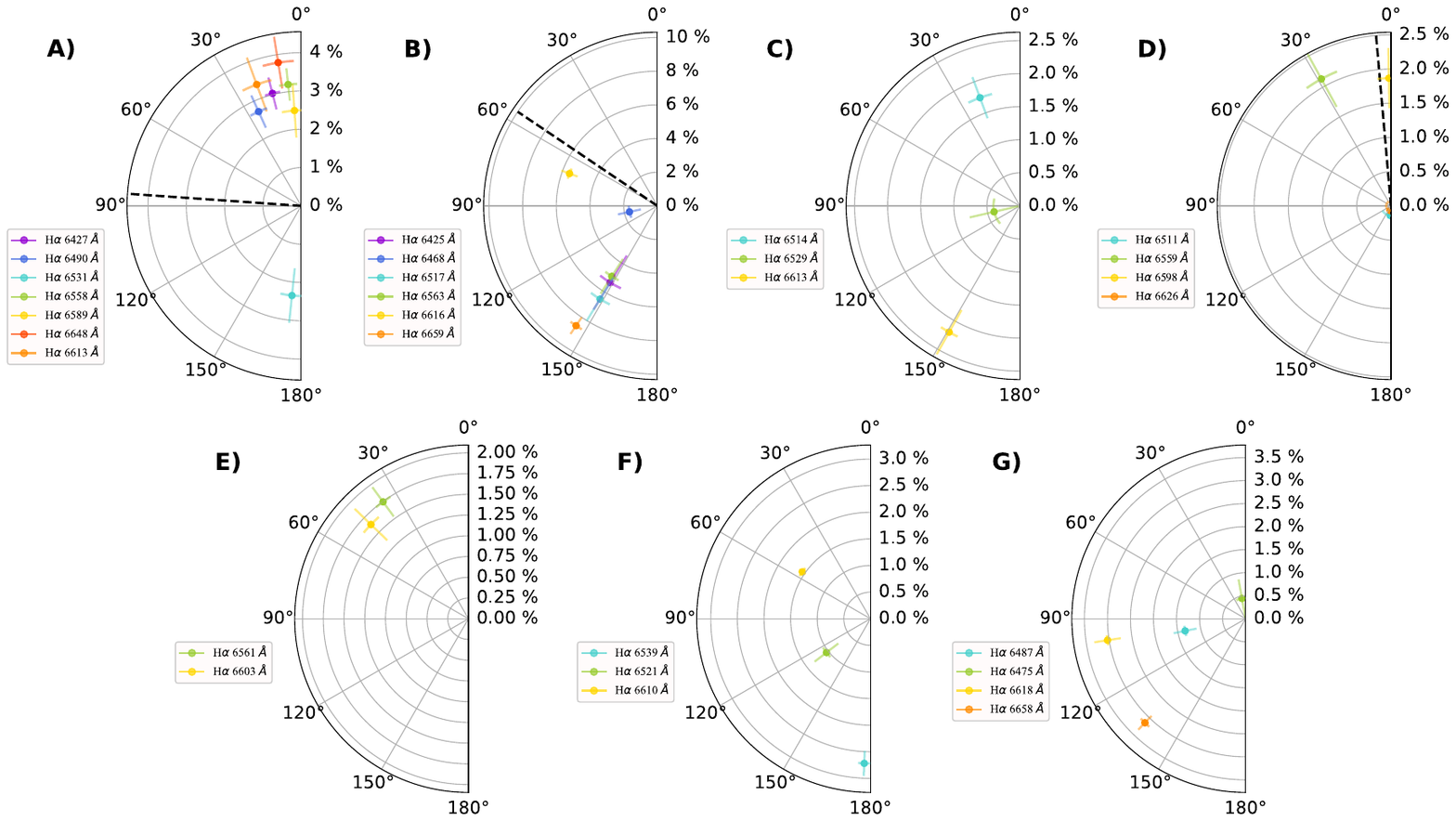}
    \caption{Protractor diagram of the polarization degree $P$ as a function of the polarization angle $\theta$ for each broad H$\alpha$ component, color-coded as in the spectral decomposition. The polarization angle is measured relative to the celestial north axis, and the radial distance represents $P$. The black dashed line indicates the radio axis when available: 3C~227 (PA $\approx 86^{\circ}$; \citealt{Prieto_1993}), PKS~0235 (PA $\approx 56^{\circ}$ estimated from the 1.5~GHz image; \citealt{Downes_1986}), and SDSS~015530 (ENLR PA $\sim5^{\circ}$ used as a proxy for the AGN axis; \citealt{Husemann_2013}). Top row: 3C~227, PKS~0235, SDSS~001224, SDSS~015530. Bottom row: SDSS~094603, SDSS~111916, SDSS~153636.}
    \label{fig:p_theta_1}
\end{figure*}

The protractor diagram of SDSS~153636 is shown in panel G of Fig.~\ref{fig:p_theta_1}. No reliable information on the radio axis is available in the literature. Nevertheless, the diagram reveals a structured distribution of the H$\alpha$ components in polarization angle. The blueshifted H$\alpha\lambda$6487 component and the redshifted H$\alpha\lambda$6618 are quasi-aligned, suggesting a common scattering geometry. Conversely, the blueshifted H$\alpha\lambda$6475 and the redshifted H$\alpha\lambda$6658 have nearly perpendicular angles, indicating distinct scattering regions. Overall, the rotation of the polarization angle across the H$\alpha$ profile reveals a complex scattering structure within the BLR, possibly probing the launch of a disk wind from the equatorial plane toward the polar direction.

\section{Discussion}
\label{discussion}

We have detected intrinsically polarized, broad, asymmetric H$\alpha$ lines in all sources of the sample. We now analyze these lines in the context of the BLR geometry and location. To do so, we present the polarizations properties as a function of the line width of individual broad H$\alpha$ components. This approach allows us to distinguish between components consistent with a Keplerian disk and those which are associated with a more polar, disk-wind structure. 

\begin{figure*}[ht]
    \centering
    \begin{minipage}[c]{1\textwidth}
        \centering
        \includegraphics[width=0.96\linewidth, trim={1.5cm 11cm 2.5cm 2cm}, clip]{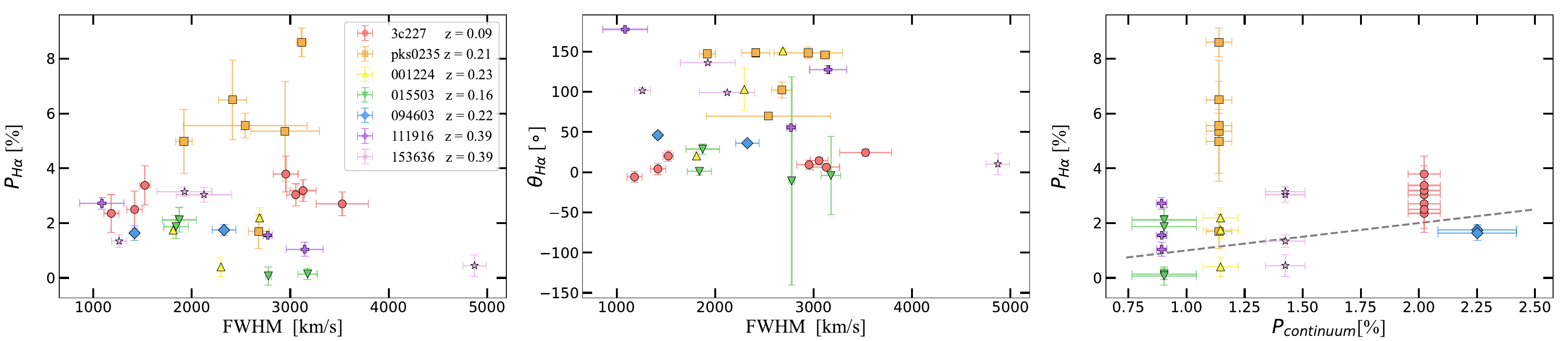} 
    \end{minipage}
    \par\medskip
    \begin{minipage}[c]{1\textwidth}
        \centering
        \includegraphics[width=0.64\linewidth, trim={1.5cm 11cm 21cm 2cm}, clip]{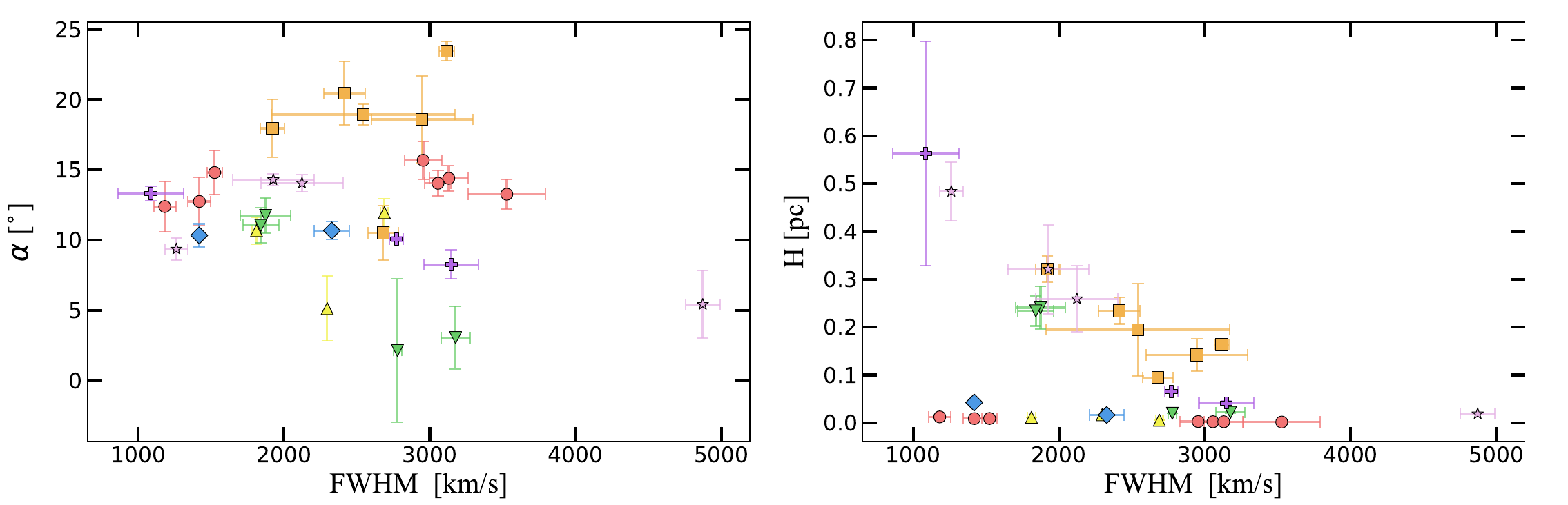}
    \end{minipage}
    \caption{Geometrical constraints on the BLR inferred from the polarization of the broad H$\alpha$ components. Top row: polarization degree $P_{H\alpha}$ and polarization angle $\theta_{H\alpha}$ as a function of the inferred FWHM, and $P_{H\alpha}$ as a function of the continuum polarization $P_{\rm continuum}$. The gray dashed line indicates $P_{H\alpha}=P_{\rm continuum}$. Bottom row: effective scattering angle $\alpha$, and BLR height as a function of the FWHM.}
    \label{fig:disussion}
\end{figure*}

The first panel of Fig.~\ref{fig:disussion} shows the polarization degree of the broad H$\alpha$ components as a function of the FWHM. Several objects (e.g., 3C~227 and PKS~0235) display highly polarized components across a wide velocity range, indicating efficient scattering throughout the BLR. Other sources show radial trends: SDSS~111916 exhibits a gradual decrease of $P_{\rm H\alpha}$ with FWHM, while SDSS~153636 presents a rise followed by a decrease toward the higher FWHM. SDSS~015503 shows two FWHM groups with different polarization levels, suggesting a transition in the dominant scattering geometry. In contrast, SDSS~094603 displays nearly constant polarization for its two components. Overall, these behaviors indicate a clear evidence for a radial stratification in the scattering geometry of the BLR.

The second panel shows the polarization angle $\theta_{\rm H\alpha}$ as a function of FWHM. In several objects (e.g., 3C~227 or SDSS~094603) the polarization angles remain relatively coherent across the BLR, indicating a dominant scattering plane. In others, significant rotations of $\theta_{\rm H\alpha}$ are observed (e.g., SDSS~001224, SDSS~111916, SDSS~153636), revealing the presence of multiple scattering geometries. PKS~0235 provides a clear example where most components share a common angle while two components deviate significantly, suggesting additional scattering contributions, possibly associated with outflowing material. These variations demonstrate that while a preferred polarization angles is often observed (indicating a dominant scattering geometry) the presence of significant rotations (up to perpendicularity) reveals the existence of additional scattering region with different orientations. The BLR scattering geometry is therefore not strictly planar, and require a vertically extended, three-dimensional structure such as disk-atmosphere or disk-wind structure.

The third panel compares the polarization degree of the broad H$\alpha$ components with that of the continuum. Three regimes are observed. In some objects (3C~227, PKS~0235, SDSS~111916) the broad H$\alpha$ components are more polarized than the continuum, implying that line emission experiences more efficient scattering. In others (SDSS~015503, SDSS~001224, SDSS~153636) the polarization levels are comparable, suggesting that the continuum and BLR emission are scattered under similar conditions. Finally, SDSS~094603 shows line polarization lower than that of the continuum, consistent with a BLR that is less affected by scattering or more directly viewed. These results indicate that the BLR and continuum do not systematically sample the same scattering regions.

The fourth panel presents the effective scattering angle $\alpha_{\rm H\alpha}$ derived from the observed line polarization using the single-scattering approximation  $P_{H_{\rm H\alpha}} = (1 - cos^{2}(\alpha_{\rm H\alpha}) )/ (1 + cos^{2}(\alpha_{\rm H\alpha}))$. The angle $\alpha_{\rm H\alpha}$ should be interpreted as a relative scattering angle rather than a direct geometrical parameter. Some objects (e.g., 3C~227, SDSS~094603) show nearly constant scattering angles across the BLR, indicating a dominant scattering geometry. Others display clear radial trends: in SDSS~111916 and SDSS~015503, broader H$\alpha$ components (larger FWHM) are associated with smaller inferred scattering angles, suggesting that scattering becomes dominated by a different region or geometry, possibly related to outflows. SDSS~153636 shows a peak at intermediate FWHM, indicating a preferred scattering region.

Finally, the last panel shows the inferred BLR height $H$ as a function of the line width FWHM, derived from $\tan(\alpha)=H/R$. Here, $R$ is the emission radius derived from $R = {M_{BH}G} / (f.FWHM_{H\alpha_{broad}}^{2})$, where $M_{BH}$ is the black hole mass taken from the literature (see Appendix.~\ref{observations:sample}), $G$ the gravitational constant, and $f$ a scaling factor. When the virial product is derived using the FWHM, a geometric factor $f=0.75$ naturally results for a spherical BLR with randomly oriented clouds \citep{Netzer_1990, Kaspi_2000, Wandel_1999, Onken_2004}. We find that the broad H$\alpha$ components follow two distinct behaviors. In the first group (3C~227, SDSS~094603, SDSS~001224), the BLR height remains approximately constant ($H < 0.05$ pc) over the full FWHM range. We note that these sources show single-peaked broad H$\alpha$ profiles. This behavior is consistent with a geometrically thick BLR characterized by a roughly constant scale height, suggesting that the vertical structure is largely decoupled from the kinematics traced by the line width. In contrast, the broad H$\alpha$ components of the second group follow an approximately linear relation, implying a roughly constant $H/R$ ratio and therefore a fixed opening angle across the BLR. The height of the BLR decreases with increasing velocity, indicating that higher-velocity components originate from regions closer to the equatorial plane, while lower-velocity components arise from geometrically thicker regions. This trend naturally points to a radially stratified BLR structure, in which the vertical extent decreases toward the inner regions. We also note that these sources show strongly asymmetric line profiles, often with pronounced broad red or blue wings. Although the individual AGNs occupy different positions in this diagram, indicating source-to-source variations, the global picture that emerges is consistent with a disk-wind geometry in which a flattened, disk-like inner BLR transitions into a more vertically extended structure at larger radii, possibly associated with outflowing material launched at intermediate angles between the equatorial plane and the polar axis.

\section{Conclusions}
\label{Conclusions}

We have obtained and analyzed optical spectropolarimetric observations of seven AGNs using the VLT/FORS2 between March 2012 and January 2013. The sample was selected based on the presence of broad and asymmetric H$\alpha$ emission line profiles in the total flux spectra, in some case showing pronounced asymmetries or double-peaked structures commonly interpreted as signatures of complex kinematics and geometries within the BLR. The primary objective of this work was to investigate whether these unusual H$\alpha$ profiles exhibit distinct polarimetric signatures, and to asses what constraints spectropolarimetry can impose on the geometry and scattering mechanisms within the BLR. To do so, we performed a spectral decomposition of the H$\alpha$ emission line for each source, modeling both narrow and broad component using Lorentzian function. This approach allowed us to isolated the intrinsic polarimetric properties, degree and angle of polarization, of the individual broad H$\alpha$ components. By combining these results with estimates of their FWHMs, we explored variations of polarization across the BLR and how different emitting regions sample the same or different scattering geometries. Below, we summarize the main polarimetric and geometric properties inferred for each AGNs :

\begin{itemize}

\item[-] 3C~227: The H$\alpha$ profile requires seven broad components. These components show relatively high polarization ($P\sim2-4\%$) and polarization angles nearly perpendicular to the radio axis, indicating dominant polar scattering. The polarization properties remain relatively stable across the BLR, suggesting a coherent, flattened scattering plane.

\item[-] PKS~0235: The double-peaked H$\alpha$ profile requires six broad components. Most of them show high polarization and angles perpendicular to the radio axis, consistent with dominant polar scattering. Two components exhibit different polarization angles, indicating additional scattering geometries and supporting a disk-wind structure.

\item[-] SDSS~001224: Three broad components reproduce the H$\alpha$ asymmetry. Their polarization angles differ strongly, revealing a rotation of $\theta$ with velocity and indicating multiple scattering geometries, consistent with emission originating from different regions of a disk-borne wind.

\item[-] SDSS~015503: Four broad components reveal two polarimetric populations. Two components show low polarization consistent with direct emission, while two others show higher polarization and different angles, suggesting distinct scattering regimes and a radially stratified BLR.

\item[-] SDSS~094603: Only two broad components are required to reproduce the profile. Their similar polarization degree and angle indicate a common scattering geometry and suggest a relatively coherent, equatorial BLR structure.

\item[-] SDSS~111916: Three broad components probe different BLR FWHM and show strong variations in polarization properties. The inner component is consistent with equatorial scattering, while outer components show stronger scattering signatures, indicating radial, elevated stratification of the scattering regions.

\item[-] SDSS~153636: Four broad components reveal strong variations of polarization degree and angle across the H$\alpha$ profile. The progressive rotation of $\theta$ suggests multiple scattering geometries and supports a structured BLR associated with a disk outflow.

\end{itemize}

In summary, this complete analysis of the sample provide a model-independent view to investigate the inner geometry of AGNs, and shows that the different broad H$\alpha$ components generally do not share a single scattering geometry. The combined variations in polarization degree and angle with FWHMs suggest an extended and stratified BLR, in which several scattering regions coexist. In many sources, these properties are consistent with the presence of a wind originating from the accretion disk, gradually connecting the inner equatorial regions to more polar structures. Although several theoretical scenarios have been proposed to explain asymmetric or doublet-peaked H$\alpha$ profiles, the available data do not allow for a clear distinction between these models for each individual source. The lack of additional information, particularly on the radio axis for certain objects and limited spectral resolution are important limitations of this study. None of the seven AGNs in our sample exhibits the horizontal S-like shape rotation of the polarization angle as a function of the radial velocity, a spectropolarimetric signature commonly observed in the majority of AGNs, for example, in the sample of \citet{Afanasiev_2019}. This absence is even more remarkable that this signature is predicted by both the standard equatorial scattering model and the off-axis illumination model proposed by \citet{Goosmann_2014}. Nevertheless, this work highlights the strong potential of high spectral resolution optical polarimetry as a geometric diagnostic tool to probe the internal geometry of AGNs, enabling an understanding of the structure of the BLR and scattering properties of the BLR.

\section{Data availability}
Table~D1 is only available in electronic form at the CDS via anonymous ftp to cdsarc.u-strasbg.fr (130.79.128.5) or via http://cdsweb.u-strasbg.fr/cgi-bin/qcat?J/A+A/.

\begin{acknowledgements}
The authors would like to acknowledge the support of the CNES, the CNRS, the University of Strasbourg, the PNHE and the PNCG. FM and JB acknowledge financial support from the french national space agency (CNES) and the Centre national de la recherche scientifique (CNRS)

\end{acknowledgements}

\bibliographystyle{aa}
\bibliography{biblio}

\begin{appendix}

\section{Observations}
\label{observations}

\subsection{Observation and data reduction}
\label{observations:data_reduction}

Spectropolarimetric observations were conducted between March 2012 and January 2013 at the European Southern Observatory (ESO) using the FOcal Reducer and low dispersion Spectrograph 2 (FORS2) mounted on the 8.2-m Antu Telescope (UT1) of the Very Large Telescope (VLT) at Paranal Observatory (Program ID ID 089.B-0969(A) ; PI C. M. Gaskell). A total of 7.3 hours of observations were obtained in service mode under good sky conditions with a fraction of lunar illumination less than 0.5 and an airmass of less than 1.3. 

FORS2 was used in the dual-beam polarimetric mode which employs a Wollaston prism and a rotating half-wave plate (HWP) to measure the linear polarization. Incoming light rays pass through the HWP before being split by the Wollaston prism into two orthogonally polarized beams: the ordinary (o-ray) and extraordinary (e-ray) rays. The HWP was rotated at four angles : 0$^{\circ}$, 22.5$^{\circ}$, 45$^{\circ}$ and 67.5$^{\circ}$ to fully characterize the linear polarization and correct for a possible differential transmission of the o- and e-rays.  The observational characteristics are summarized in Table~\ref{tableaux_observation}. The spectra of 3C~227, PKS0235+023, SDSSJ001224-102226, SDSSJ015530-085704, and SDSSJ153636+044127 were acquired using the 300V grism with the filter GG435 to cover the 4450-8650~\AA\ spectral range. A 1.3" slit was used, providing a minimum spectral resolution of R$\sim$340 at the central wavelength of 5849~\AA. The spectra of SDSSJ094603+013923 and SDSSJ111916+110107 were secured using the 300I grism with the filter OG590 to cover the 6000-11000~\AA\ spectral range. The 1.3" slit provided a minimum spectral resolution of R$\sim$500 at the central wavelength of 8575~\AA. The slit was aligned with the parallactic angle, which is given in  Table~\ref{tableaux_observation}.  The spatial scale on the detector was 0$\farcs$25 per pixel with a 2$\times$2 binning. For each HWP angle, the exposure time was 294~s. The sequence of four exposures was repeated twice for  a total exposure time of 2352~s per target.

\begin{table*}[h!]
    \centering

    \caption{Log of the VLT/FORS2 observations.}
    \resizebox{\textwidth}{!}{
    \begin{tabular}{c c c c c c c c}
        \hline\\[-9pt] 
        \large\textbf{Source} 
        & \large\textbf{Date} 
        & \multicolumn{2}{c}{\large\textbf{J2000}} 
        & \large\textbf{z} 
        & \large\textbf{PA slit [\degr]} 
        & \large\textbf{Seeing ["]} 
        & \large\textbf{Airmass} \\
        & & \large\textbf{RA} & \large\textbf{DEC} & & & & \\
        \hline
        
        3C~227
        &14 Jan 2013
        &09 47 45.10
        &+07 25 20.0
        &0.0860
        &6
        & {0.6}
        & {1.2}
        \\
        
        PKS0235 +023
        &21 Oct 2012
        &02 38 32.68
        &+02 33 50.2
        &0.2075
        &165
        & {0.6}
        & {1.2}
        \\
        
        SDSS J001224 -102226
        &15 Jul 2012
        &00 12 24.08
        &-10 22 25.8
        &0.2284
        &54
        & {1.0}
        & {1.1}
        \\
        
        SDSS J015530 -085704
        &22 Jul 2012
        &01 55 30.01
        &-08 57 04.0
        &0.1646
        &55
        & {1.3}
        & {1.1}
        \\
        
        SDSS J094603 +013923
        &14 Jan 2013
        &09 46 04.11
        &+01 39 25.0
        &0.2196
        &33
        & {0.9}
        & {1.1}
        \\
        
        SDSS J111916 +110107
        &31 Mar 2012
        &11 19 16.13
        &+11 01 07.1 
        &0.3933
        &160
        & {1.1}
        & {1.3}
        \\
        
        SDSS J153636 +044127
        &31 Mar 2012
        &15 36 36.26 
        &+04 41 28.0 
        &0.3890 
        &34
        & {1.3}
        & {1.2}
        \\
        
        \hline
    \end{tabular}}
    \tablefoot{Columns (1) – source name; (2) – observation date; (3) – Ra(J2000); (4) – Dec(J2000); (5) – redshift ; (6) – slit position angle; (7) – average seeing ; (8) – average airmass.}
    \label{tableaux_observation}
\end{table*}

The raw data were first corrected for cosmic rays using a Python implementation of the \textsc{lacosmic} algorithm \citep{Dokkum_2001,Dokkum_2012}. Data reduction was performed using the ESO FORS2 pipeline to obtain images with two-dimensional spectra rectified and calibrated in wavelength. The one-dimensional spectra were extracted using a 6\arcsec -long subslit centered on the nucleus. The sky spectrum was estimated from two well-separated regions in the 20\arcsec -long slit and subtracted from the nucleus spectrum. Then, the normalized Stokes parameters $q(\lambda)$ and $u(\lambda)$, the linear polarization degree $p(\lambda)$, the polarization position angle $\theta(\lambda)$, and the total flux density $F(\lambda)$ were computed from the ordinary and extraordinary spectra according to standard recipes \citep{Marin2025,Bagnulo2009}.  The direct spectrum $F(\lambda)$ was corrected for extinction and calibrated in flux using a master response curve.  The spectra were corrected for the rotator angle and for the retarder plate zero angle provided in the FORS2 manual, so that $\theta(\lambda)$ = 0$\degr$ (90$\degr$) corresponds to the north (east) direction.  Uncertainties were estimated by propagating the photon and readout noises. Finally, the polarization degree $P$ and the polarization angle $\theta$ were computed using the standard equations: $P$ = $\sqrt{q^{2} + u^{2}}$ and $\theta$ = $\frac{1}{2} \arctan \left( \frac{u}{q} \right)$.

\subsection{The sample}
\label{observations:sample}

Our sample consists of seven type 1 AGNs exhibiting extremely asymmetric and broad Balmer emission lines and sufficiently high brightness (V$\leqslant$17.5), making them well studied for details spectropolarimetric analysis. Throughout this work, we adopt a flat $\Lambda$CDM cosmology with $H_{0}$=70km.s$^{-1}$Mpc$^{-1}$, $\Omega_{\Lambda}$=0.7, $\Omega_{m}$=0.3.

\subsubsection{3C 227}
3C 227 is a famous Broad-Line Radio Galaxy (BLRG) located at a redshift z$\sim$0.086 (comoving distance 362 Mpc). This object presents one of the broadest and most luminous Extended Emission Line Region (EELR), covering a scale of $\sim$200 kpc and showing partial alignment with the radio structure \citep{Prieto_1993}. Geometrical and kinematic simulations of the EELR in 3C~227 showed that the ionized gas can be modeled as a bicone with a large opening angle of 120$^{\circ}$, inclined at $40^{\circ}$ to the plane of the sky \citep{Prieto_1993}. Because its H$\alpha$ emission line has a luminosity of $\sim 7.5 \times 10^{42}$ erg.s$^{-1}$, the authors concluded that 3C 227 could be a radio-loud quasar observed close to the edge of the ionization cone, with a mass of the central black hole estimated at $log M_{BH} = 7.13~M_{\odot}$ \citep{Mezcua_2011, Prieto_1993}. 3C~227 was observed in polarization in 1994 by \citet{Marshall_1999} using the Keck telescope. H$\alpha$ observations reveal a $\sim$ 20$^{\circ}$ difference in the polarization angle between the line and the continuum. The authors theorized the existence of several lines of sight into the nuclear continuum source and the BLR, spatially separated by a few tens of parsecs, so different scatterers are responsible for the two polarizations. The observed rotation of the polarization angle across the broad H$\alpha$ line suggests that the BLR clouds are dominated by organized streaming motions, possibly in equatorial orbits around the central engine, rather than chaotic cloud velocities. These results support a viewing angle close to the limit of the dusty torus, allowing a partial view of the nucleus. Finally, \citet{Gezari_2007} have studied 3C 227 from a series of spectra obtained between 1990 and 2004; they showed that the mean H$\alpha$ profile has a pronounced blue central peak at v = -1500 km~s$^{-1}$, with a broad base extending to blue and red wavelengths, producing an asymmetric structure. During this period, the excess of the flux in the blue peak flux decreased while the red-to-blue flux ratio, measured from high variability in the root mean square (RMS) spectrum, on the red side (1700 to 3400~km~s$^{-1}$) and on the blue side (-1200 to -4400~km~s$^{-1}$), increased. The evolution of the H$\alpha$ profile is modeled by two Gaussian components shifting towards the red between 1990 and 2000. The authors stated that such drift is not consistent with a single disk in rotation.  

\subsubsection{PKS 0235+023}
PKS 0235+023 is a broad line Seyfert 1 galaxy at a redshift z$\sim$ 0.209 comoving distance 852 Mpc), which hosts a black hole whose mass was estimated at $log M_{BH} = 8.805~M_{\odot}$, together with a nuclear luminosity of $log(L_{B}) = 44.1$ erg.s$^{-1}$ \citep{Wu_2004, Sikora_2007}. Interestingly, the AGN displays an asymmetric H$\alpha$ profile with double peaks, also studied by \citet{Gezari_2007}. The mean spectrum of the profile shows a blue peak at V(B) = -4700~km~s$^{-1}$ and a red peak at V(R)=+2700~km~s$^{-1}$, producing a strongly asymmetric profile. Their analysis of the RMS spectrum reveals notable variability in the wings of the profile. During the period between 1991 and 2005, the $\frac{F(R)}{F(B)}$ ratio fell by 50\% indicating a diminution of the red side of the profile while the blue peak gradually shifted from -3600~km~s$^{-1}$ to -5000~km~s$^{-1}$. If a circular disk model correctly reproduces the profile in 1991 \citep{Eracleous_1994}, it fails to reproduce the subsequent profiles observed thereafter. 

\subsubsection{SDSS 001224-102226}
SDSS 0012-1022 (FBQS J0012-1022; WISEA J001224.02-102226.5) is a quasar with a luminosity of $\log L_{5100\,\text{\AA}} =$ 44.43 erg.s$^{-1}$ at 5100~\AA \, located at redshift z$\sim$ 0.22 (comoving distance 894 Mpc). Its central black hole mass is estimated at $log M_{BH}$ = 8.584 $M_{\odot}$ \citep{Wu_2004}. This object shows a notable evolution of the asymmetric H$\alpha$ profile between the SDSS DR7 spectra identified by \citet{Tsalmantza_2011} and the observations made during the period 2011-2012 by \citet{Decarli_2013}, see also \citet{Zhang2025}. The H$\alpha$ line profile shows a diminution of the structure present in the red wing, while the blue wing becomes more pronounced and shifts toward shorter wavelengths \citep{Decarli_2013}.  

\subsubsection{SDSS 015530-085704}
SDSS 0155-0857 (WISEA J015530.02-085704.2) is a quasar located at a redshift z $\sim$ 0.1648 or 679 Mpc \citep{Runnoe_2015}. It displays a blue asymmetric H$\beta$ line profile with a black hole mass estimated at $log M_{BH}$ = 8.85 $M_{\odot}$ \citep{Du_2018, Kim_2010}. \citet{Tsalmantza_2011} reported slight redshift of the H$\alpha$ line ($\sim$ 1550~km~s$^{-1}$) with a slight asymmetry in the profile, something that is even more visible in the H$\beta$ line profile ($\sim$ 2200~km~s$^{-1}$).  

\subsubsection{SDSS 094603+013923}
SDSS 0946+0139 (WISEA J094603.94+013923.7) is a quasar at z$\sim$0.2196 (comoving distance 893 Mpc) with a black hole mass estimated at $log M_{BH}$ = 7.91 $M_{\odot}$ \citep{Shin_2021}, displaying complex emission lines. The luminosity of the H$\alpha$ line, measured from SDSS spectra, is about 10$\times$10$^{41} erg.s^{-1}$ \citep{Decarli_2013}. This object presents a particularly unusual H$\beta$ line profile, characterized by a strong redshifted peak and marked blue asymmetry, as reported by \citet{Eracleous_2012} and \citet{Du_2018}. This profile was interpreted as consistent with a supermassive black hole binary scenario, as proposed by \citet{Runnoe_2017}.

\subsubsection{SDSS 111916+110107}
SDSS 1119+1101 (WISEA J111916.12+110107.1) is a broad-line quasar located at a redshift z $\sim$ 0.3936 or 1529 Mpc \citep{Runnoe_2025}, with very little known information. The black hole mass of this AGN was estimated to be $log(M_{BH}) = 8.69~M_{\odot}$ \citep{Li_2008}. It is present in the ROSAT (1RXS) and 2MASS catalogs, and in the W2R (WISE+2MASS+RASS) sample \citep{Edelson2012}. It has a log$_{10}$ bolometric luminosity of 45.61 $\pm$ 0.09 erg s$^{-1}$ and its log$_{10}$ BLR radius was estimated to be 1.88 $\pm$ 0.02 light-days \citep{Baldini2015}.

\subsubsection{SDSS 153616+044127}
SDSS 1536+0441 (SDSS J153636.22+044127.0) hosts a radio-quiet quasar at redshift z~=~0.389 or a comoving distance of 1513 Mpc \citep{Decarli_2009, Decarli_2013} with a radio luminosity $L_{R} = \nu L_{\nu}$ of $5.2\times10^{40} erg~s^{-1}$ at 8.5 GHz and a bolometric luminosity estimated at  $L_{bol} = 1.5 \times 10^{46} erg.s^{-1}$ \citep{Wrobel_2009, Wrobel_2010}. This quasar is known for its broad, double-peaked emission lines in the optical spectrum, with each peak separated by around 3500 $km~s^{-1}$ \citep{Boroson_2009}. This unusual spectral morphology has led the authors to identify SDSS~1536+0441 as a candidate for hosting a sub-parsec binary SMBH system with masses of $10^{7.3}M_{\odot}$ and $10^{8.9}M_{\odot}$, with a separation of $\sim$ 0.1 pc and an orbital period of $\sim$ 100 years. However, several later studies have challenged this hypothesis. \citet{Chornock_2009} proposed that this object is instead an atypical double-peaked emitter, where the velocity separation of the broad lines arises from rotational and relativistic motions within an accretion disk around a single SMBH.

\section{Optical spectropolarimetric observations}
\label{app:data}

\begin{figure}[H]
    \centering
    \begin{minipage}[c]{0.45\textwidth}
        \centering
        \includegraphics[width=\linewidth, trim={3cm 5.5cm 7cm 12cm}, clip]{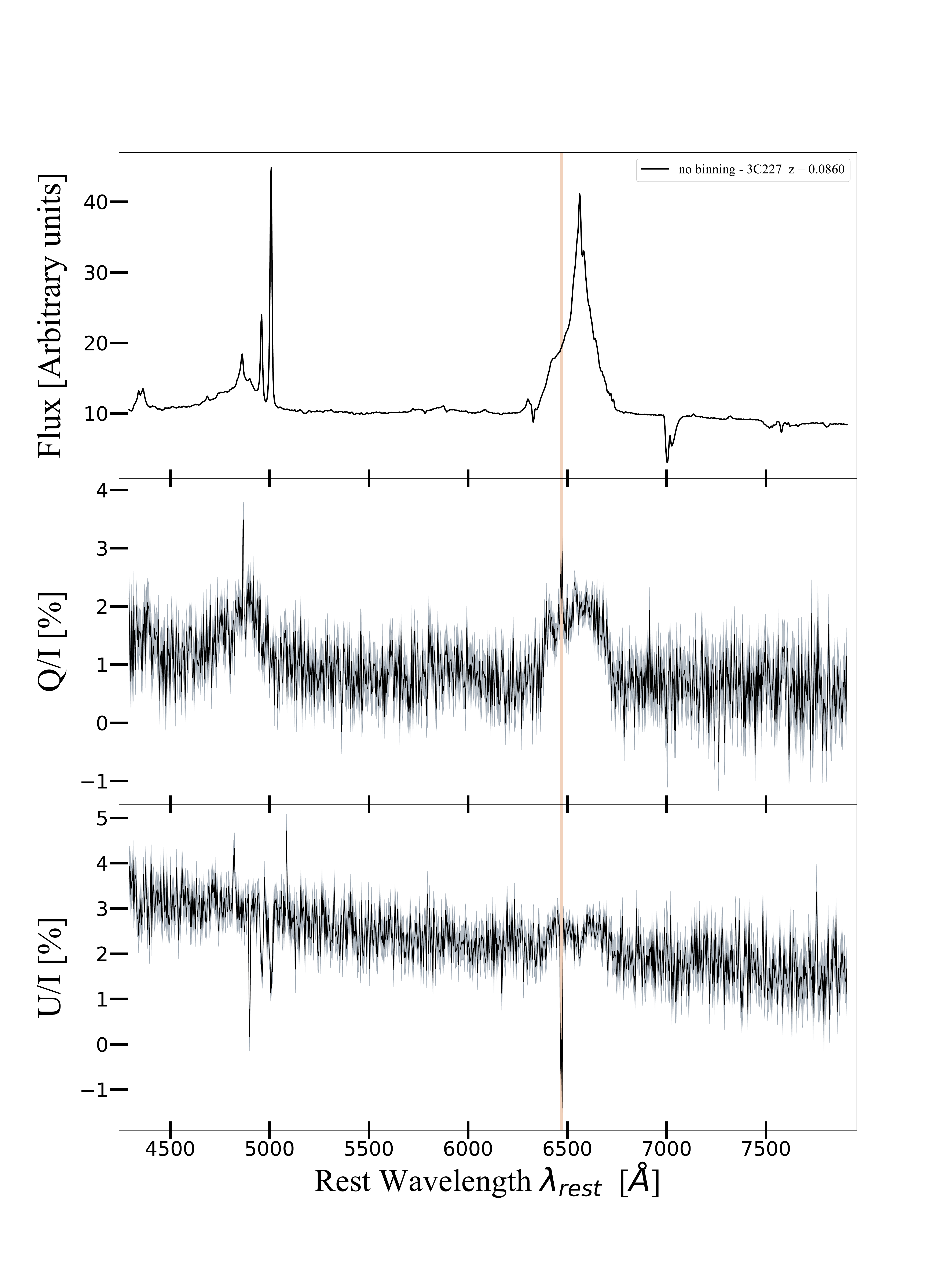}
    \end{minipage}
    \hspace{0.5pt}
    \begin{minipage}[c]{0.45\textwidth}
        \centering
        \includegraphics[width=\linewidth, trim={3cm 5.5cm 7cm 12cm}, clip]{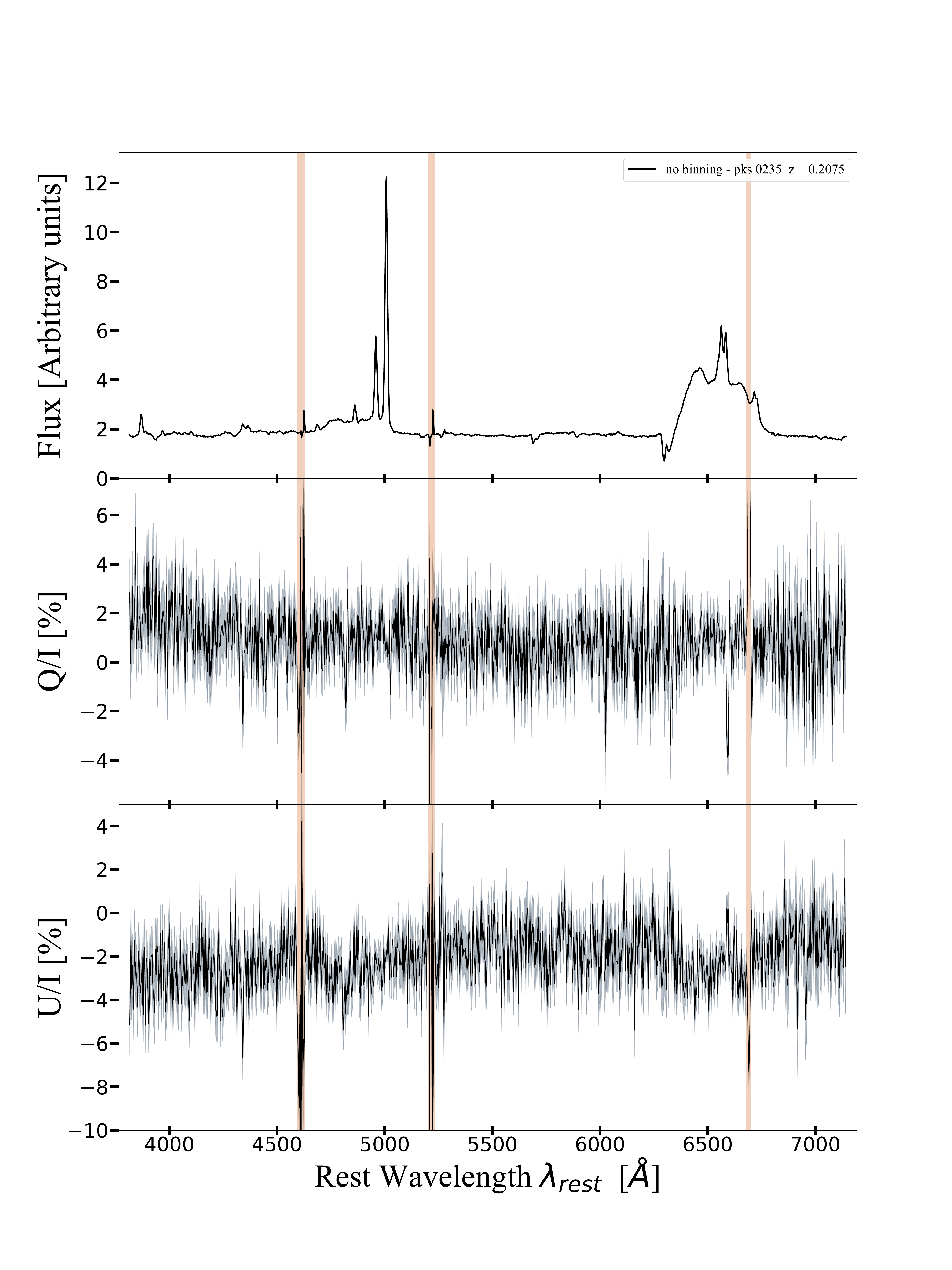}
    \end{minipage}
        \vspace{0.5cm}
    \begin{minipage}[c]{0.45\textwidth}
        \centering
        \includegraphics[width=\linewidth, trim={1.5cm 5.5cm 7cm 11.5cm}, clip]{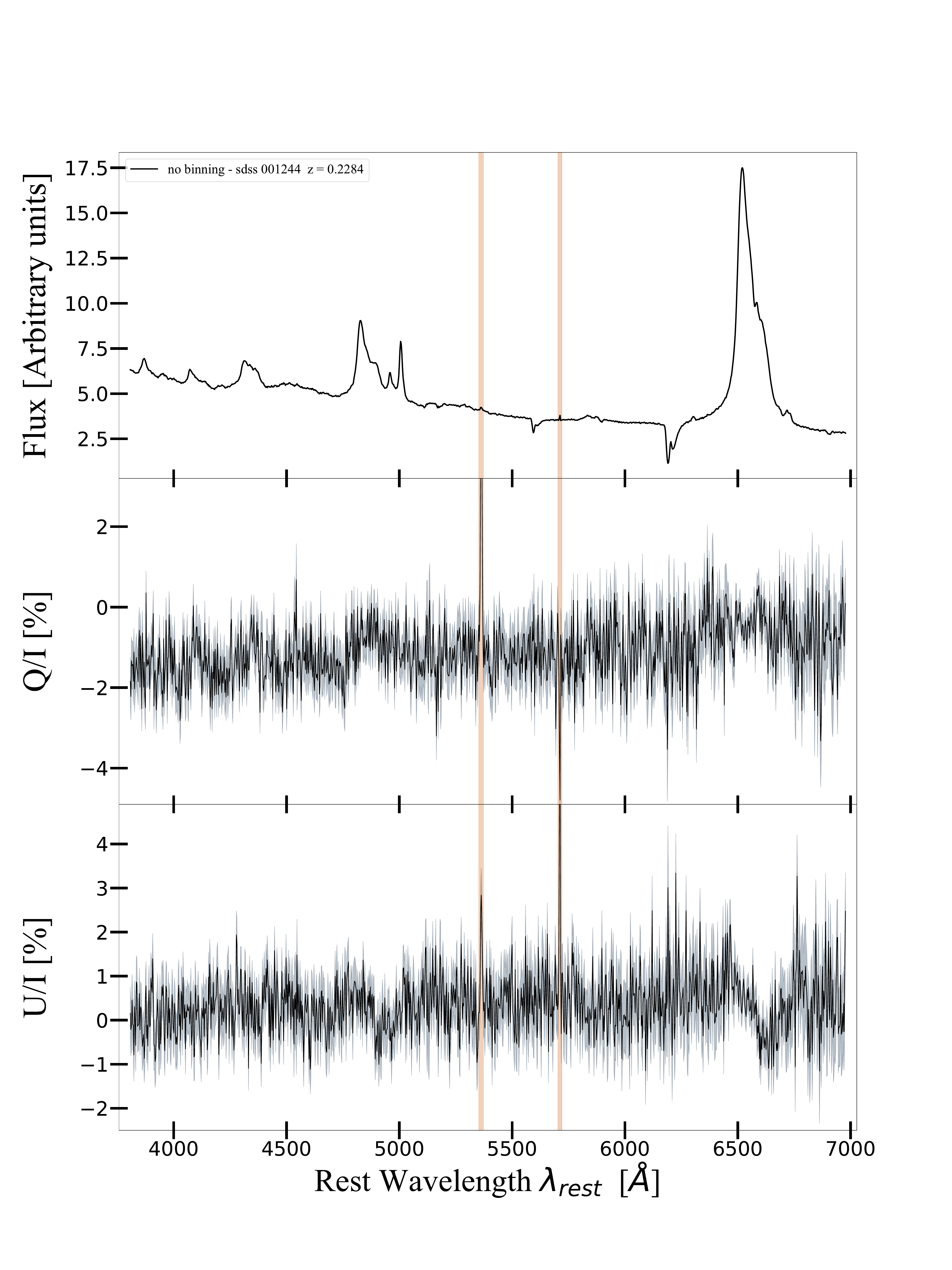}
    \end{minipage}
    \hspace{0.5pt}
    \begin{minipage}[c]{0.45\textwidth}
        \centering
        \includegraphics[width=\linewidth, trim={1.5cm 5.5cm 7cm 11.5cm}, clip]{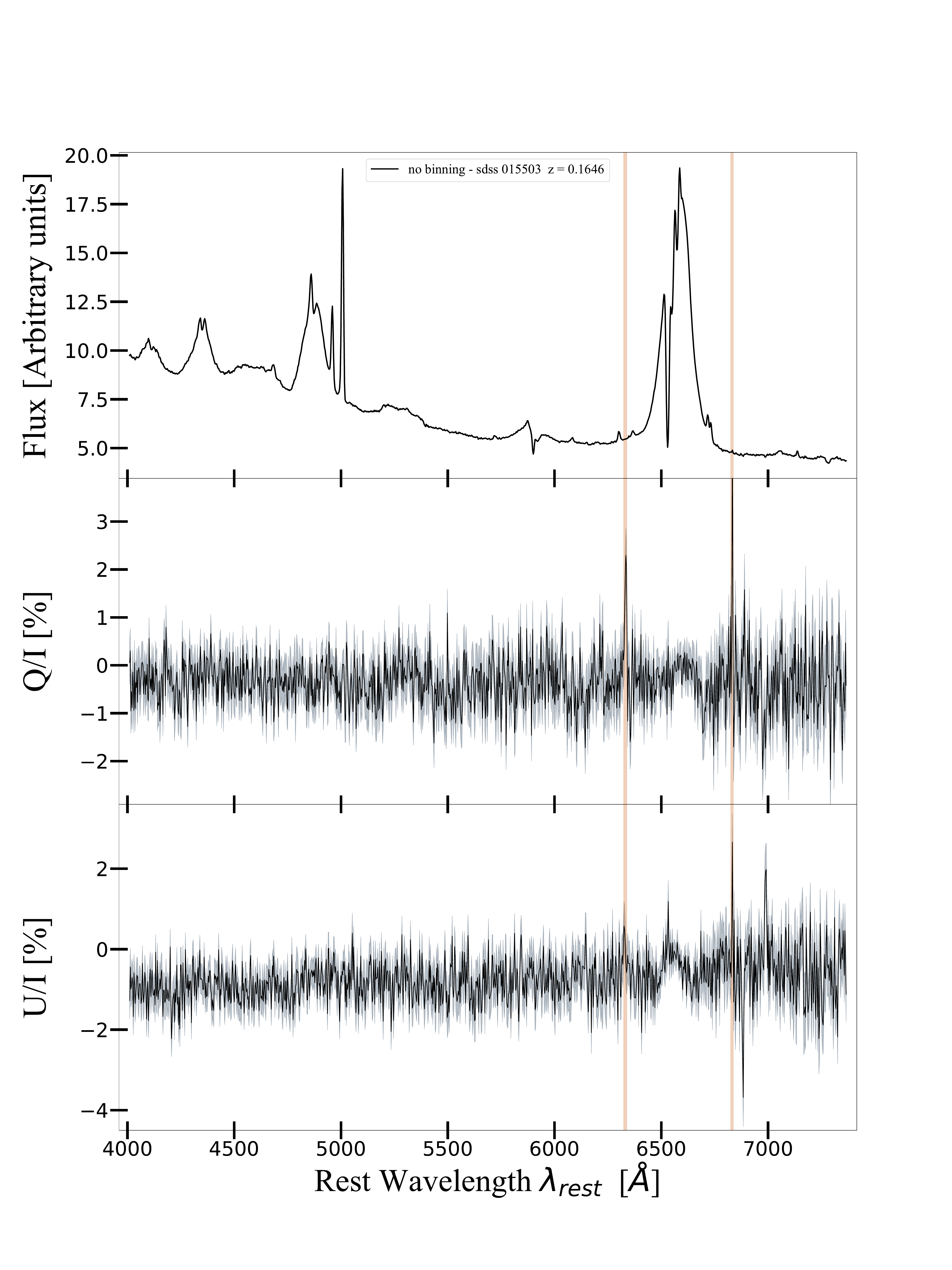}
    \end{minipage}

    \caption{Spectra obtained for each observation with FORS2/VLT. Each panel present, from top to bottom, the total flux, and the fraction of Stokes parameters $q=Q/I$ and $u=U/I$ as a function of the wavelength. Top:  3C~227 on the left figure and PKS~0235 for the right figure. Spectral range identified as an artifact and which has been removed are the flowing: for 3C~227, 6462-6478~\AA \ ; for PKS~0235, 4594-4629~\AA, 6676-6698~\AA, 5200-5230~\AA. Bottom: SDSS~001224, 5352-5373\AA , 5703-5720~\AA \ on the left and SDSS~015530, 6825-6837~\AA \ on the right.}
    \label{fig:annexe1_data_1}
\end{figure}

\begin{figure}[H]
    \centering
    \begin{minipage}[c]{0.45\textwidth}
        \centering
        \includegraphics[width=\linewidth, trim={1.5cm 6cm 7cm 12cm}, clip]{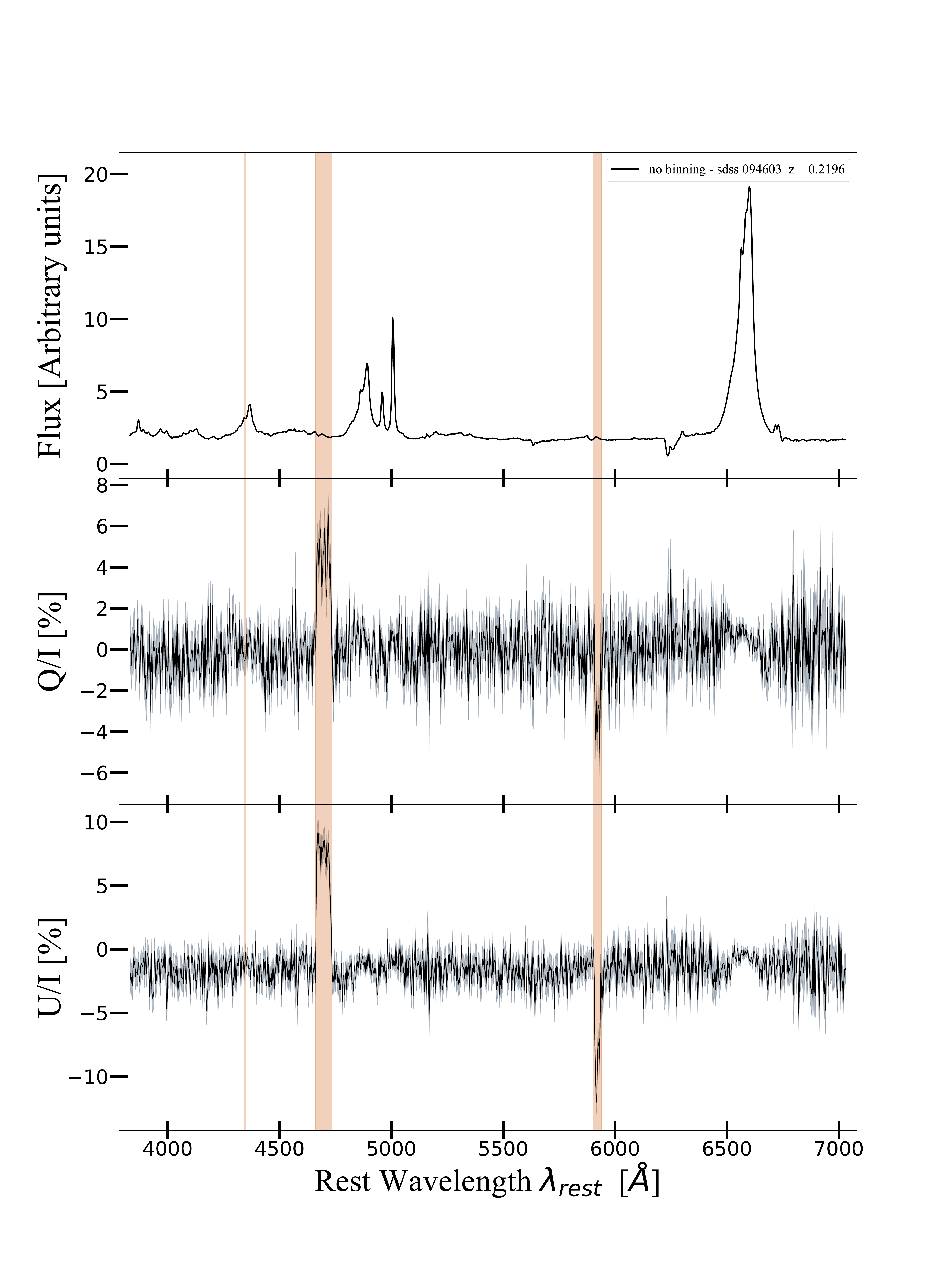}
    \end{minipage}
    \hspace{0.5pt}
    \begin{minipage}[c]{0.45\textwidth}
        \centering
        \includegraphics[width=\linewidth, trim={1.5cm 6cm 7cm 12cm}, clip]{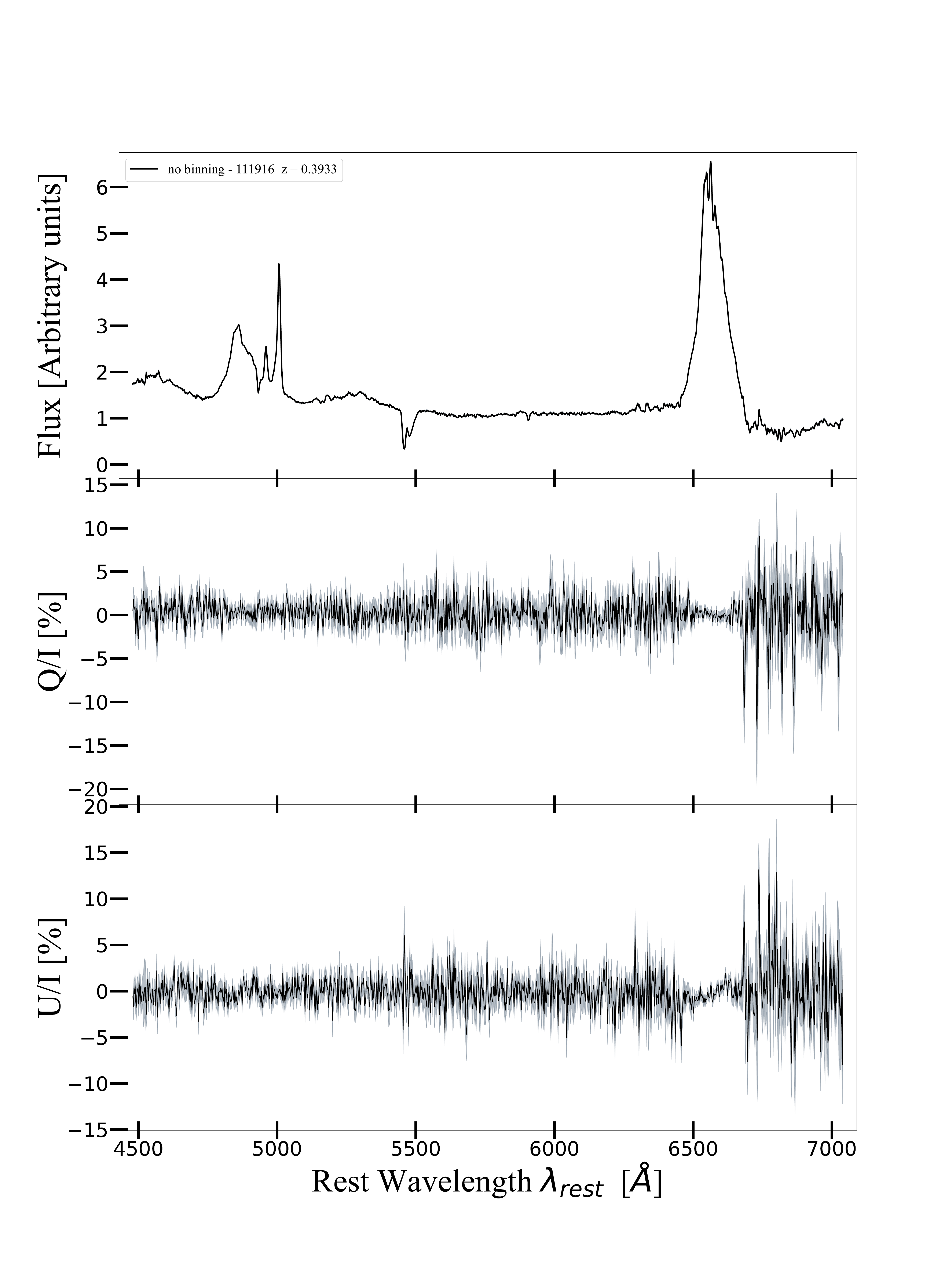}
    \end{minipage}

    \vspace{0.5cm}
    \begin{minipage}[c]{0.45\textwidth}
        \centering
        \includegraphics[width=\linewidth, trim={1.5cm 6cm 7cm 11cm}, clip]{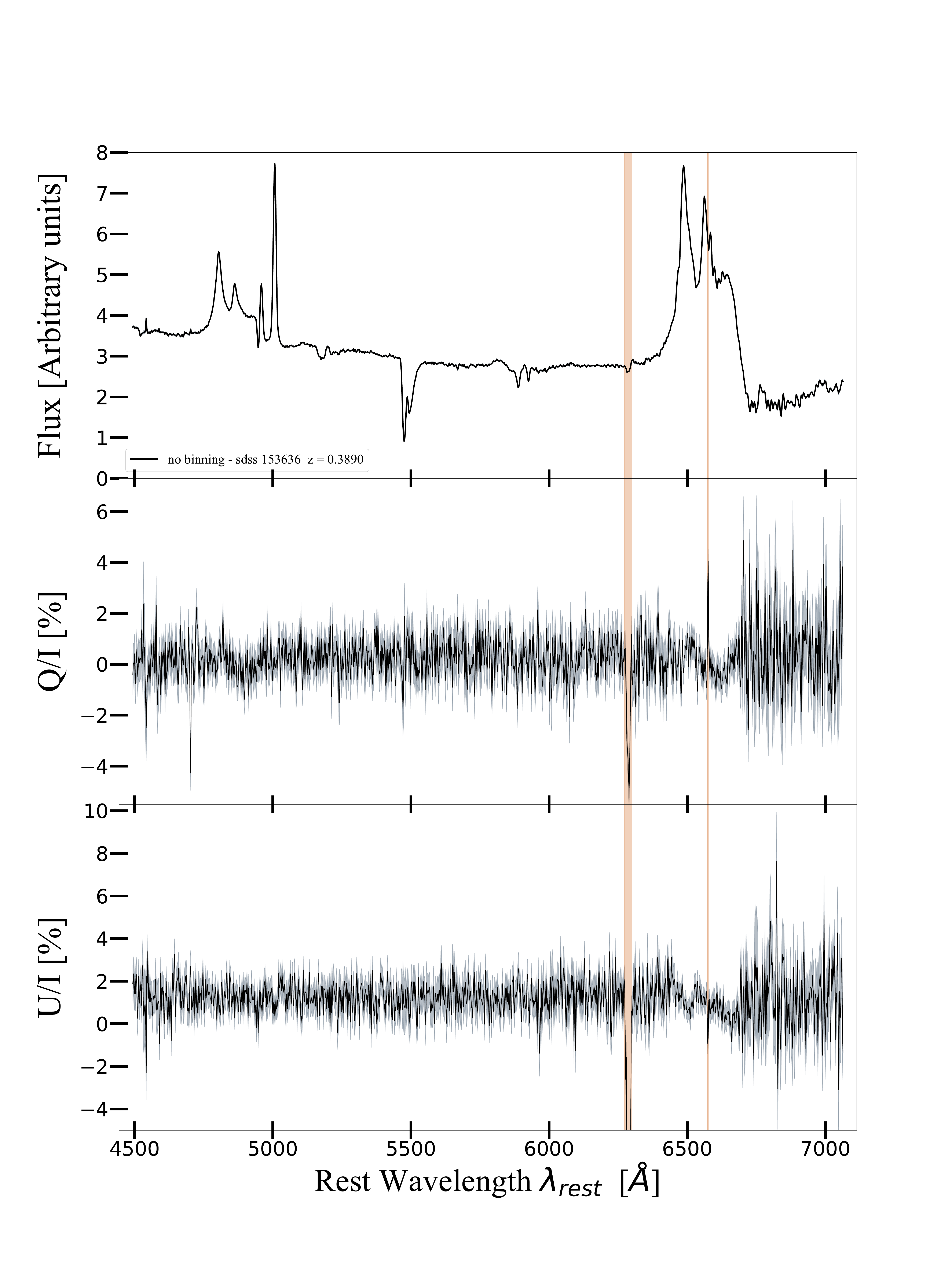}
    \end{minipage}
    \hspace{0.5pt}
    \begin{minipage}[c]{0.45\textwidth}
        \centering
    \end{minipage}

    \caption{Same as \ref{fig:annexe1_data_1}, top: SDSS~094603 (left), 4660-4731~\AA, 5902-5940\AA \ and SDSS~111916 (right), bottom: SDSS~153636, 6272-6300~\AA , 6573-6579~\AA.}
    \label{fig:annexe1_data_2}
\end{figure}

Despite a careful data reduction, residual instrumental and observational artifacts were identified in the spectral region corresponding to the H$\alpha$ profile. As shown in Fig.~\ref{fig:annexe1_data_1} and \ref{fig:annexe1_data_2}, these artifacts were manually removed and then replaced by a local interpolation, when possible. The affected wavelength ranges are small, a few tens of angstroms at most, and have been carefully corrected in order to not alter the shape of the profile nor the outcomes of our analysis. We made sure that these adjustments have no impact on the general structure of the H$\alpha$ profile.

\onecolumn

\twocolumn

\section{Fitting results}

As a first step, the underlying continuum was modeled using a power-law function, estimated from line-free regions surrounding the local H$\alpha$ complex profile. This continuum component was then subtracted from the observed spectra. The residual total intensity spectra were subsequently decomposed by fitting all detected narrow emission lines ($i_{narrow}$) from [O~I]$\lambda$6300 to [S~II]$\lambda$6730 (rest frame) using Lorentzian profiles. The central wavelengths of the narrow lines were identified from local maxima in the continuum-subtracted spectra and fixed during the fitting procedure, while their widths and amplitudes were left free. To reproduce the asymmetric H$\alpha$ profiles observed in each source, we then added one or more broad H$\alpha$ components ($i_{broad}$), also modeled with Lorentzian functions. In contrast to the narrow lines, the central wavelengths of the broad components were treated as free parameters in order to capture velocity shifts and asymmetries in the line profiles. The number of broad components was kept to the minimum required to adequately reproduce the observed spectra. Each narrow emission line is thus characterized by two free parameters, its full width at half maximum (FWHM) and its amplitude ($\theta_{narrow}\equiv FWHM_{i_{narrow}}, I_{i_{narrow}}$), while each broad H$\alpha$ component is described by three free parameters: FWHM, amplitude, and central wavelength ($\theta_{broad}\equiv\,FWHM_{i_{broad}}, I_{i_{broad}}, x_{0, i_{broad}}$). The best-fit model parameters were obtained through a $\chi^{2}$ minimization procedure. The resulting best-fit spectral models are shown as red solid lines in the top panels of each figure presented in Section~\ref{Analysis}. The individual broad H$\alpha$ components are highlighted in color, while the narrow emission lines are shown in black.

~\

\begin{figure}[!t]
    \centering
    \includegraphics[width=0.90\linewidth, trim={2cm 11cm 6.4cm 15cm}, clip]{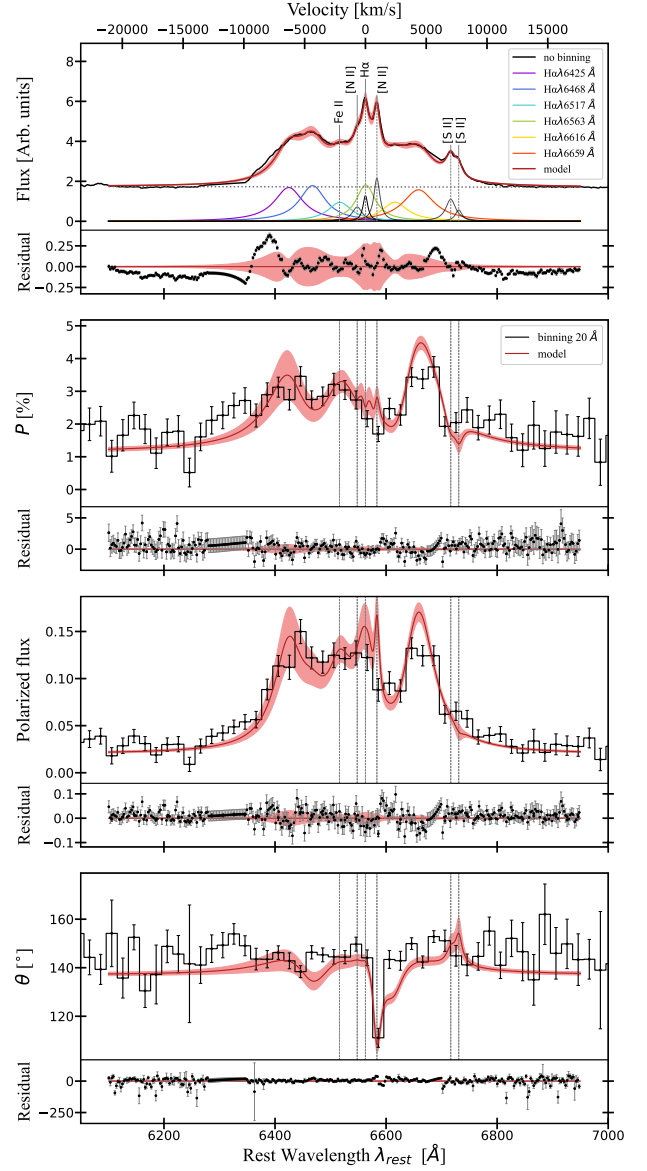}
    \caption{Same as \ref{Fig3C227_spectraldecomposition} but for PKS~0235}
    \label{FigPKS0235_spectraldecomposition}
\end{figure}

\begin{figure}[!t]
    \centering
    \includegraphics[width=0.90\linewidth, trim={2cm 11cm 6.4cm 15cm}, clip]{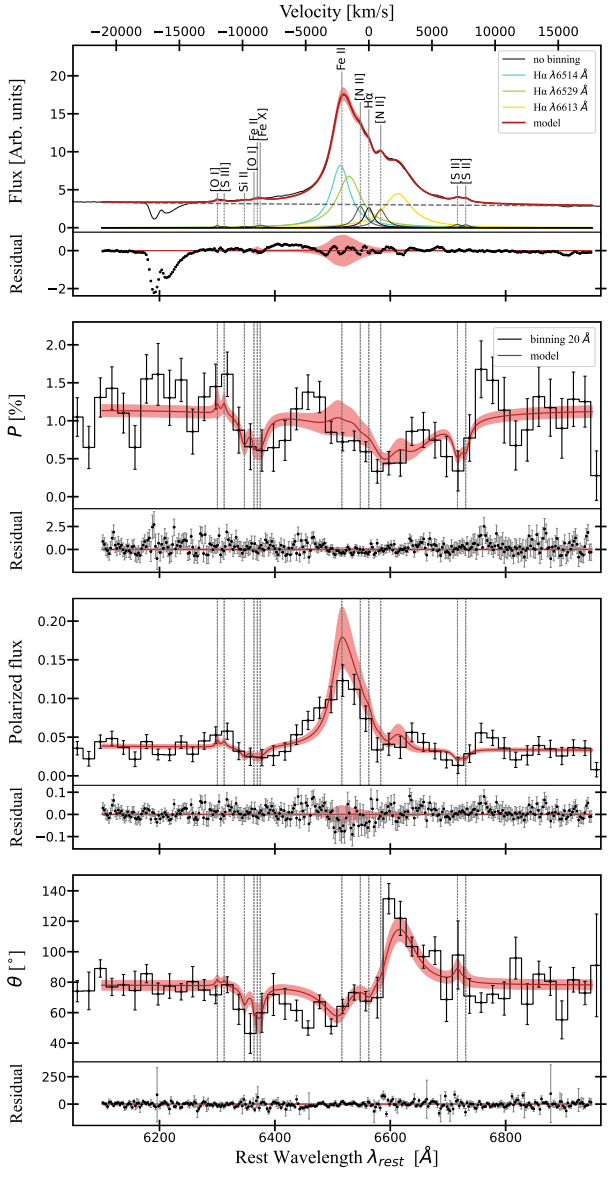}
    \caption{Same as \ref{Fig3C227_spectraldecomposition} but for SDSS~001244}
    \label{Fig001244_spectraldecomposition}
\end{figure}

\begin{figure}[!t]
    \centering
    \includegraphics[width=0.90\linewidth, trim={2cm 11cm 6.4cm 15cm}, clip]{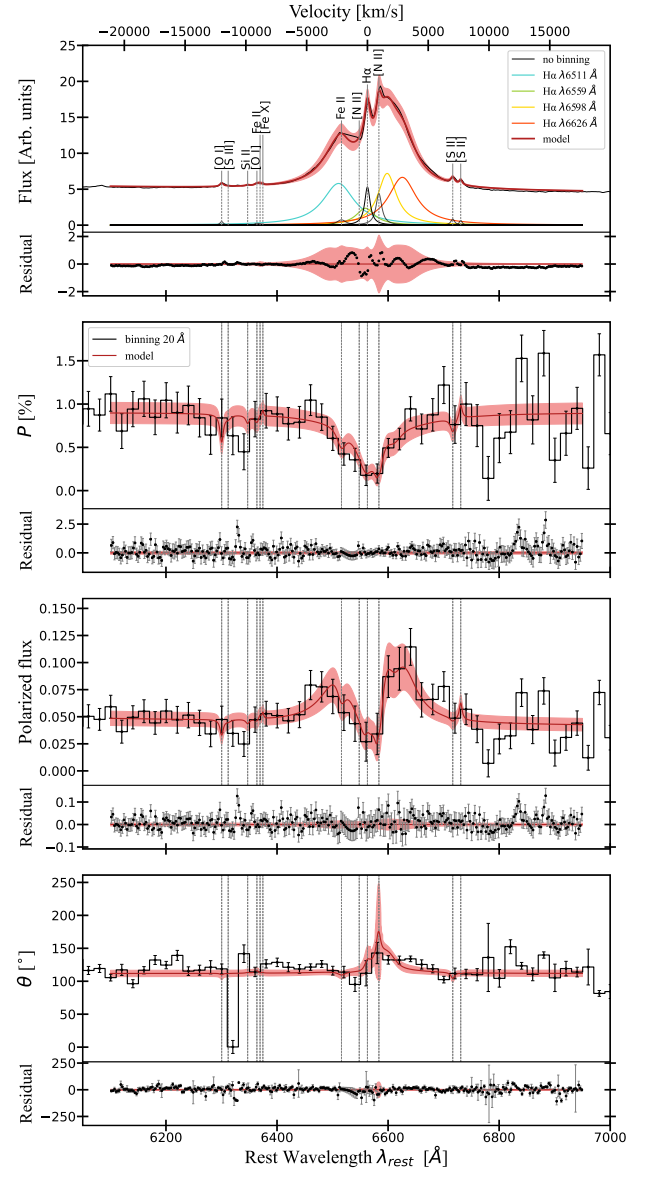}
    \caption{Same as \ref{Fig3C227_spectraldecomposition} but for SDSS~015503}
    \label{Fig015503_spectraldecomposition}
\end{figure}

\begin{figure}[!t]
    \centering
    \includegraphics[width=0.90\linewidth, trim={2cm 11cm 6.4cm 15cm}, clip]{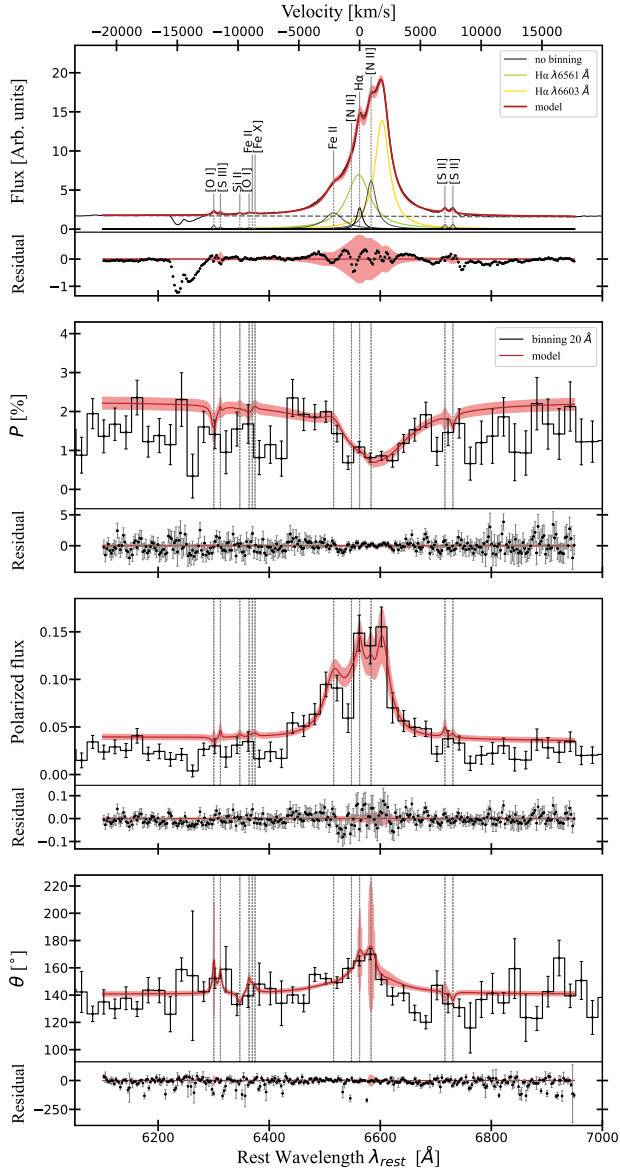}
    \caption{Same as \ref{Fig3C227_spectraldecomposition} but for SDSS~094603}
    \label{Fig094603_spectraldecomposition}
\end{figure}

\begin{figure}[!t]
    \centering
    \includegraphics[width=0.90\linewidth, trim={2cm 11cm 6.4cm 15cm}, clip]{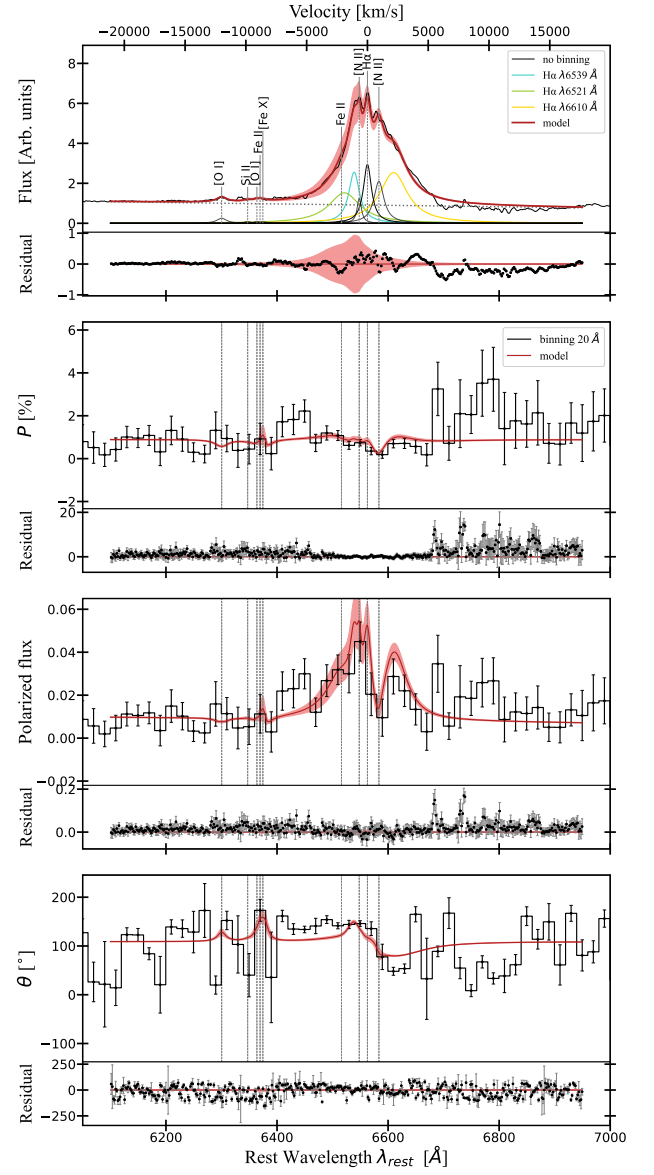}
    \caption{Same as \ref{Fig3C227_spectraldecomposition} but for SDSS~111916}
    \label{Fig111916_spectraldecomposition}
\end{figure}

\begin{figure}[!t]
    \centering
    \includegraphics[width=0.90\linewidth, trim={2cm 11cm 6.4cm 15cm}, clip]{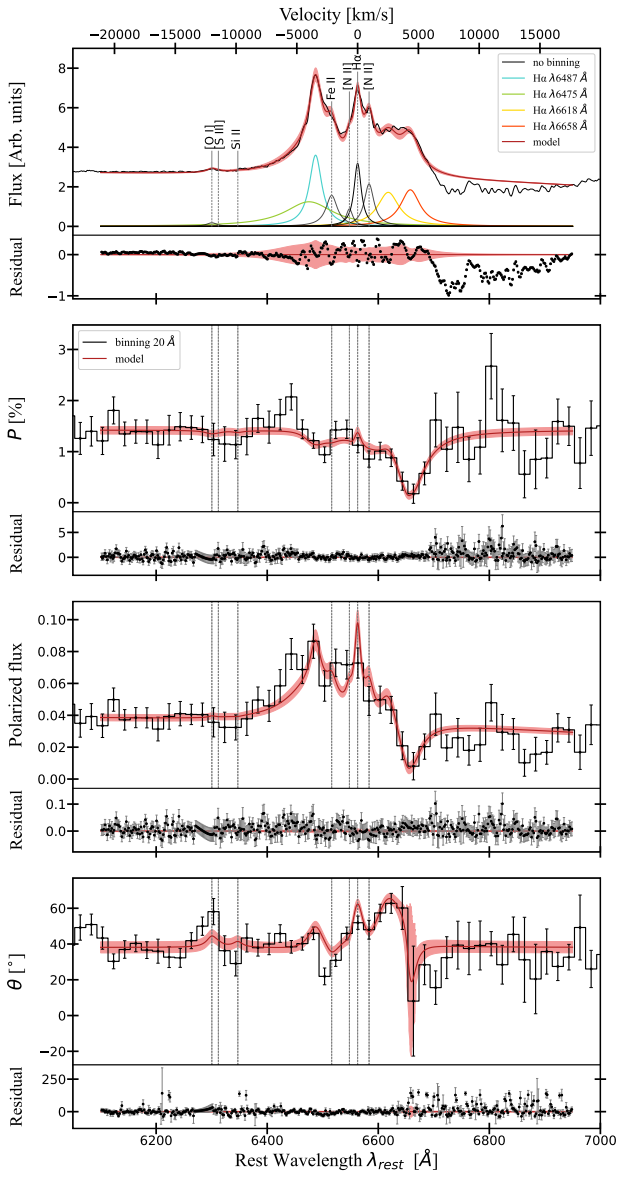}
    \caption{Same as \ref{Fig3C227_spectraldecomposition} but for SDSS~153636}
    \label{Fig153636_spectraldecomposition}
\end{figure}

Our next step was to investigate the polarimetric response of the spectral model. Using the Stokes formalism \citep{Stokes_1852}, we constructed a polarized model by assigning a wavelength-independent polarization state to each spectral component identified in the total flux fit (continuum, narrow emission lines, and broad emission lines). For each spectral component $i$, modeled either as a power-law or a Lorentzian profile with total intensity $I_{i}(\lambda)$, we defined the associated Stokes parameters as $Q_{i}(\lambda) = q_{i} \times I_{i}(\lambda)$ and $U_{i}(\lambda) = u_{i} \times I_{i}(\lambda)$, where $q_{i}$ and $u_{i}$ are dimensionless fractional polarization coefficients constrained to the interval $[-1,1]$. These coefficients encode both the polarization degree and polarization angle of each individual emission component, and are assumed to be constant with wavelength across the H$\alpha$ spectral region at first order. Such hypothesis comes from the fact that BLR clouds are ionized and thus, scattering is mostly grey (Thomson scattering). The total Stokes parameters of the model are obtained by linear superposition of all spectral components, $Q(\lambda) = \sum_{i} Q_{i}(\lambda)$ and $U(\lambda) = \sum_{i} U_{i}(\lambda)$, while the total intensity $I_{i}(\lambda)$ is fixed by the best-fit spectral model. From the modeled Stokes parameters, we computed the polarization degree $P$ and polarization angle $\theta$ following the formulas already presented in Appendix.~\ref{observations:data_reduction}. The polarization model therefore introduces two free parameters ($q_{i}$ and $u_{i}$) for each emission component contributing to the H$\alpha$ profile. The model parameters were optimized through a $\chi^{2}$ minimization procedure by fitting simultaneously the observed polarization degree and polarization angle. Since the Stokes parameters are additive, the best-fit values obtained from the polarization degree and angle were combined to derive the final estimates of $q_{i}$ and $u_{i}$ for each spectral component. The resulting best-fit polarized model is shown as the solid red line in Figs.~\ref{Fig3C227_spectraldecomposition}, \ref{FigPKS0235_spectraldecomposition}, \ref{Fig001244_spectraldecomposition}, \ref{Fig015503_spectraldecomposition}, \ref{Fig094603_spectraldecomposition},  \ref{Fig111916_spectraldecomposition}, and \ref{Fig153636_spectraldecomposition}. We summarize the best-fit spectral and polarization parameters for all broad and narrow emission components in the online supplementary material (Table D.1).

\onecolumn
\section{Best fit spectral and polarimetric parameters of the decomposition of the H{\ensuremath{\alpha}} profile}
\label{app:param_model}

\small{
\setlength{\tabcolsep}{4pt} 
\begin{longtable}{p{1.5cm} | c c c c c c c c}
\caption{Parameters and best-fit values for narrow and broad component of the total flux and polarization model for each source of our sample.}\\
        \hline\\[-10pt] 
        &  & \textbf{Intensity} & \textbf{FWHM} & \textbf{$x_{0}$} & \textbf{q} & \textbf{u} & \textbf{$P$} & \textbf{$\theta$} \\
        & & \textbf{[arb. units]} & \textbf{[km~s$^{-1}$]} & \textbf{[\AA]} &  & & \textbf{[\%]} & \textbf{[\%]}  \\
        \hline\\[-1pt] 
    \endfirsthead
        \hline\\[-10pt] 
        &  & \textbf{Intensity} & \textbf{FWHM} & \textbf{$x_{0}$} & \textbf{q} & \textbf{u}& \textbf{$P$} & \textbf{$\theta$} \\
        & & \textbf{[arb. units]} & \textbf{[km~s$^{-1}$]} & \textbf{[\AA]} &  & & \textbf{[\%]} & \textbf{[\%]}  \\
        \hline\\[-1pt] 
    \endhead

        \multirow{16}{*}{\textbf{3C227}} 
        & continuum & - & - & -                                   & 0.004 $\pm$0.001   & 0.020 $\pm$ 0.001 & 2.0 $\pm$ 0.1 & 40 $\pm$ 1 \\
        & [O~I]     & 1.24 $\pm$ 0.09 & 479 $\pm$ 7     & 6300.30 & 0.015 $\pm$ 0.011  & -0.004 $\pm$ 0.007  & 1.5 $\pm$ 1.1 & 173 $\pm$ 14  \\
        & [S~III]   & 0.13 $\pm$ 0.05 & 369 $\pm$ 52    & 6312.06 & -0.049 $\pm$ 0.024 & 0.037 $\pm$ 0.040 & 6.2 $\pm$ 3.1 & 71 $\pm$ 16 \\
        & [Fe~X]    & 0.46$\pm$ 0.02 & 1212 $\pm$ 97    & 6374.51 & 0.0247 $\pm$ 0.015 & -0.002 $\pm$ 0.011 & 2.5 $\pm$ 1.5 & 178 $\pm$ 13 \\
        & Fe II     & 0.26 $\pm$ 0.42 & 972 $\pm$ 987  & 6516.08 & 0.142 $\pm$ 0.019 & -0.002 $\pm$ 0.064 & 14.2 $\pm$ 1.9 & 180 $\pm$ 13\\
        & [N~II]    & 2.82 $\pm$ 0.16 & 814 $\pm$ 102  & 6548.05 & 0.023 $\pm$ 0.001 & -0.006 $\pm$ 0.006 & 2.4 $\pm$ 1.0 & 173 $\pm$ 8\\
        & H$\alpha$ & 9.57 $\pm$ 0.09 & 561 $\pm$ 12    & 6562.82 & -0.008 $\pm$ 0.007 & -0.007 $\pm$ 0.005 & 0.7 $\pm$ 0.5 & 131 $\pm$ 33\\
        & [N~II]    & 2.85 $\pm$ 0.22 & 478 $\pm$ 34    & 6583.46 & -0.027 $\pm$ 0.011 & 0.007 $\pm$ 0.011 & 2.8 $\pm$ 1.1 & 83 $\pm$ 11\\
        & [S II]    & 0.95 $\pm$ 0.03 & 798 $\pm$ 74    & 6716.44 & 0.017 $\pm$ 0.010 & 0.015 $\pm$ 0.016 & 2.2 $\pm$ 1.3 & 21 $\pm$  18 \\
        & [S II]    & 0.47 $\pm$ 0.02 & 221 $\pm$ 21    & 6730.81 & 0.016 $\pm$ 0.016 & 0.011 $\pm$ 0.019  & 1.9 $\pm$ 1.7 & 18 $\pm$  27\\
        & H$\alpha$   & 5.07 $\pm$ 0.66 & 3056 $\pm$ 91 & 6427.13 $\pm$ 0.66  & 0.027 $\pm$ 0.004 & 0.015 $\pm$ 0.003   & 3.0 $\pm$ 0.4 & 14 $\pm$ 3\\
        & H$\alpha$   & 6.05 $\pm$ 1.80 & 3528 $\pm$ 264  & 6489.54 $\pm$ 1.80  & 0.018 $\pm$ 0.003 & 0.020 $\pm$ 0.005 & 2.7 $\pm$ 0.4 & 24 $\pm$ 4\\
        & H$\alpha$   & 4.20 $\pm$ 0.42 & 1182 $\pm$ 76  & 6531.06 $\pm$ 0.95  & 0.023 $\pm$ 0.007 & -0.005 $\pm$ 0.005 & 2.4 $\pm$ 0.7 & 174 $\pm$ 7\\
        & H$\alpha$   & 16.37 $\pm$ 0.25 & 3130 $\pm$ 133  & 6558.46 $\pm$ 1.97  & 0.031 $\pm$ 0.004 & 0.007 $\pm$ 0.004 & 3.2 $\pm$ 0.4 & 6 $\pm$ 4\\
        & H$\alpha$   & 5.71 $\pm$ 0.40 & 1419 $\pm$ 78  & 6589.11 $\pm$ 0.79  & 0.0247 $\pm$ 0.007 & 0.004 $\pm$ 0.006 & 2.5 $\pm$ 0.7 & 4 $\pm$ 7\\
        & H$\alpha$   & 5.18 $\pm$ 0.10 & 1524 $\pm$ 52  & 6613.09 $\pm$ 0.55  & 0.026 $\pm$ 0.007 & 0.022 $\pm$ 0.008 & 3.4 $\pm$ 0.7 & 20 $\pm$ 6\\
        & H$\alpha$   & 5.96 $\pm$ 0.13 & 2955 $\pm$ 126  & 6647.77 $\pm$ 0.39  & 0.036 $\pm$ 0.006 & 0.012 $\pm$ 0.008 & 3.8 $\pm$ 0.6 & 9 $\pm$ 5\\
        &   \\
        \hline\\[-3pt]
        \multirow{11}{*}{\textbf{PKS~0235}} 
        & continuum & - & - & -                                 & 0.001 $\pm$ 0.001  &  -0.011 $\pm$ 0.001 & 1.1 $\pm$ 0.1 & 137 $\pm$ 1 \\
        & Fe II     & 0.10 $\pm$ 0.01 & 916 $\pm$ 86  & 6516.08 & 0.001 $\pm$ 0.002  & 0.016 $\pm$ 0.002 & 1.6 $\pm$ 0.2 & 43 $\pm$ 4\\
        & [N~II]    & 0.71 $\pm$ 0.08 & 1007 $\pm$ 1  & 6548.05 & 0.006 $\pm$ 0.003  & 0.016 $\pm$ 0.003 & 1.8 $\pm$ 0.3 & 34 $\pm$ 5 \\
        & H$\alpha$ & 1.29 $\pm$ 0.18 & 489 $\pm$ 88  & 6562.82 & 0.006 $\pm$ 0.003  & 0.017 $\pm$ 0.002 & 1.8 $\pm$ 0.2 & 35 $\pm$ 4 \\
        & [N~II]    & 2.15 $\pm$ 0.19 & 607 $\pm$ 86  & 6583.46 & -0.072 $\pm$ 0.009 & 0.013 $\pm$ 0.001 & 7.3 $\pm$ 0.9 & 85 $\pm$ 0.7 \\
        & [S II]    & 1.10 $\pm$ 0.03 & 982 $\pm$ 4  & 6716.44  & 0.011 $\pm$ 0.008  & 0.016 $\pm$ 0.002 & 1.9 $\pm$ 0.5 & 27 $\pm$ 9\\
        & [S II]    & 0.56 $\pm$ 0.03 & 662 $\pm$ 26  & 6730.81 & 0.022 $\pm$ 0.011  & 0.034 $\pm$ 0.018 & 4.1 $\pm$ 1.6 & 29 $\pm$ 10\\      
        & H$\alpha$   & 1.69 $\pm$ 3.86 & 2946 $\pm$ 348  & 6424.70 $\pm$ 3.86  & 0.024 $\pm$ 0.011  & -0.048 $\pm$ 0.020 & 5.3 $\pm$ 1.8 & 146 $\pm$ 7 \\
        & H$\alpha$   & 1.78 $\pm$ 1.88 & 2679 $\pm$ 104  & 6467.57 $\pm$ 1.89  & -0.015 $\pm$ 0.006 & -0.007 $\pm$ 0.006 & 1.7 $\pm$ 0.6 & 102 $\pm$ 10 \\
        & H$\alpha$   & 0.94 $\pm$ 0.08 & 2414 $\pm$ 142  & 6516.77 $\pm$ 0.66  & 0.029 $\pm$ 0.011  & -0.058 $\pm$ 0.015 & 6.5 $\pm$ 1.4 & 148 $\pm$ 5 \\
        & H$\alpha$   & 1.82 $\pm$ 0.25 & 1921 $\pm$ 83  & 6563.42 $\pm$ 1.13   & 0.021 $\pm$ 0.007  & -0.045 $\pm$ 0.001 & 5.0 $\pm$1.1 & 147 $\pm$ 5 \\
        & H$\alpha$   & 0.96 $\pm$ 0.04 & 2542 $\pm$ 629  & 6616.02 $\pm$ 0.96  & -0.042 $\pm$ 0.003 & 0.036 $\pm$ 0.006 & 5.6 $\pm$ 0.4 & 70 $\pm$ 3 \\
        & H$\alpha$   & 1.57 $\pm$ 0.11 & 3116 $\pm$ 50  & 6658.51 $\pm$ 0.51   & 0.032 $\pm$ 0.007  & -0.080 $\pm$ 0.005 & 8.6 $\pm$ 0.5 & 146 $\pm$ 2 \\
        &   \\
        \hline\\[-3pt] 
        \multirow{16}{*}{\textbf{SDSS~0012}} 
        & continuum & - & - & -                                 & -0.011 $\pm$ 0.001  &  0.005 $\pm$ 0.001  & 1.1 $\pm$ 0.1 & 78 $\pm$ 3   \\
        & [O~I]     & 0.34 $\pm$ 0.02 & 336 $\pm$ 19  & 6300.30 & -0.018 $\pm$ 0.021 & -0.012 $\pm$ 0.007 & 2.1 $\pm$  1.7 & 107 $\pm$  17  \\
        & [S~III]   & 0.19 $\pm$ 0.03 & 404 $\pm$ 32  & 6312.06 & -0.035 $\pm$ 0.026 & -0.015 $\pm$ 0.008 & 3.8 $\pm$  2.4 & 101 $\pm$  9  \\
        & Si~II     & 0.17 $\pm$ 0.01 & 726 $\pm$ 43  & 6347.09 & 0.101 $\pm$ 0.028 & 0.009 $\pm$ 0.009 & 10.2 $\pm$  2.8 & 3 $\pm$  2 \\
        & [O~I]     & 0.11 $\pm$ 0.08 & 219 $\pm$ 135  & 6363.78 & 0.101 $\pm$ 0.027 & 0.009 $\pm$ 0.010 & 10.1 $\pm$  2.7 & 3 $\pm$  3  \\
        & Fe II     & 0.04 $\pm$ 0.19 & 458 $\pm$ 483  & 6369.47 & 0.112 $\pm$ 0.029 & 0.006 $\pm$ 0.009 & 11.2 $\pm$  2.8 & 2 $\pm$  2\\
        & [Fe~X]    & 0.33 $\pm$ 0.08 & 983 $\pm$ 225  & 6374.51 & 0.076 $\pm$ 0.025 & 0.013 $\pm$ 0.009 & 7.7 $\pm$  2.5 & 5 $\pm$  4\\
        & Fe II     & 0.05 $\pm$ 0.18 & 1012 $\pm$ 1  & 6516.08 & -0.002 $\pm$ 0.004 & -0.003 $\pm$ 0.003 & 0.3 $\pm$  0.4 & 119 $\pm$  31\\
        & [N~II]    & 2.81 $\pm$ 0.08 & 1007 $\pm$ 1  & 6548.05 & 0.006 $\pm$ 0.004 & -0.002 $\pm$ 0.004 & 0.6 $\pm$  0.4 & 173 $\pm$  20 \\
        & H$\alpha$ & 2.64 $\pm$ 0.11 & 872 $\pm$ 15  & 6562.82 & 0.009 $\pm$ 0.002 & 0.007 $\pm$ 0.006 & 1.1 $\pm$  0.4 & 18 $\pm$  12  \\
        & [N~II]    & 2.40 $\pm$ 0.04 & 1001 $\pm$ 2  & 6583.46 & 0.008 $\pm$ 0.003 & -0.001 $\pm$ 0.005 & 0.8 $\pm$  0.3 & 178 $\pm$  17 \\
        & [S II]    & 0.43 $\pm$ 0.04 & 785 $\pm$ 75  & 6716.44 & 0.033 $\pm$ 0.004 & -0.026 $\pm$ 0.006 & 4.2 $\pm$  0.5 & 161 $\pm$ 3\\
        & [S II]    & 0.39 $\pm$ 0.03 & 665 $\pm$ 91  & 6730.81 & 0.030 $\pm$ 0.006 & -0.014 $\pm$ 0.007 & 3.3 $\pm$  0.6 & 167 $\pm$ 6\\  
        & H$\alpha$   & 8.22 $\pm$ 0.57 & 1811 $\pm$ 29 & 6513.61 $\pm$ 1.23  & 0.013 $\pm$ 0.002 & 0.011 $\pm$ 0.004 & 1.7 $\pm$ 0.3 & 20 $\pm$ 6\\
        & H$\alpha$   & 6.76 $\pm$ 0.54 & 2296 $\pm$ 1  & 6528.57 $\pm$ 0.98  & -0.004 $\pm$ 0.004 & -0.002 $\pm$ 0.004 & 0.4 $\pm$ 0.4 & 103 $\pm$ 26\\
        & H$\alpha$   & 4.48 $\pm$ 0.05 & 2689 $\pm$ 25  & 6613.21 $\pm$ 0.32  & 0.011 $\pm$ 0.002 & -0.019 $\pm$ 0.004 & 2.2 $\pm$ 0.4 & 151 $\pm$ 3\\
        &   \\
        \hline\\[-3pt] 
        \multirow{10}{*}{\textbf{SDSS~0155}} 
        & continuum & - & - & -                                  &-0.007 $\pm$ 0.001  &  -0.006 $\pm$ 0.002 & 0.9 $\pm$ 0.1 & 112 $\pm$ 4   \\
        & [O~I]     & 0.50 $\pm$ 0.03 & 343 $\pm$ 30   & 6300.30 & 0.023 $\pm$ 0.009 & 0.030 $\pm$ 0.009 & 3.8 $\pm$ 0.9 & 26 $\pm$ 7  \\
        & [S~III]   & 0.06 $\pm$ 0.02 & 576 $\pm$ 169  & 6312.06 & 0.023 $\pm$ 0.008 & 0.030 $\pm$ 0.008 & 3.8 $\pm$ 0.8 & 26 $\pm$ 6 \\
        & Si~II     & 0.12 $\pm$ 0.09 & 393 $\pm$ 219  & 6347.10 & 0.039 $\pm$ 0.008 & 0.021 $\pm$ 0.005 & 3.5 $\pm$ 0.7 & 14 $\pm$ 4 \\
        & [O~I]     & 0.27 $\pm$ 0.04 & 421 $\pm$ 93   & 6363.77 & 0.032 $\pm$ 0.010 & -0.002 $\pm$ 0.008 & 3.2 $\pm$ 1.0 & 178 $\pm$ 8  \\
        & Fe II     & 0.22 $\pm$ 0.14 & 338 $\pm$ 213  & 6369.46 &-0.010 $\pm$ 0.010 & -0.003 $\pm$ 0.010 & 1.1 $\pm$ 0.9 & 98 $\pm$ 26  \\
        & [Fe~X]    & 0.15 $\pm$ 0.09 & 141 $\pm$ 145  & 6374.51 &-0.010 $\pm$ 0.009 & -0.027 $\pm$ 0.009 & 2.9 $\pm$ 0.9 & 125 $\pm$ 9 \\
        & Fe II     & 0.74 $\pm$ 0.86 & 925 $\pm$ 215  & 6516.08 & 0.018 $\pm$ 0.003 & 0.044 $\pm$ 0.004 & 3.7 $\pm$ 0.4 & 34 $\pm$ 2.2 \\
        & [N~II]    & 0.03 $\pm$ 0.05 & 248 $\pm$ 357  & 6548.05 &-0.028 $\pm$ 0.012 & -0.017 $\pm$ 0.009 & 3.3 $\pm$ 1.1 & 105 $\pm$ 9 \\
        & H$\alpha$ & 5.31 $\pm$ 0.95 & 546 $\pm$ 64   & 6562.82 & 0.009 $\pm$ 0.002 & 0.005 $\pm$ 0.003 & 1.0 $\pm$ 0.2 & 14 $\pm$  7 \\
        & [N~II]    & 4.40 $\pm$ 1.66 & 575 $\pm$ 165  & 6583.46 & 0.012 $\pm$ 0.004 & 0.020 $\pm$ 0.004 & 2.32 $\pm$ 0.4 & 30 $\pm$ 5 \\
        & [S II]    & 0.84 $\pm$ 0.07 & 376 $\pm$ 27   & 6716.44 &-0.001 $\pm$ 0.006 & 0.023 $\pm$ 0.008 & 2.3 $\pm$ 0.8 & 46 $\pm$ 8 \\
        & [S II]    & 0.66 $\pm$ 0.09 & 282 $\pm$ 48   & 6730.81 &-0.012 $\pm$ 0.009 & -0.009 $\pm$ 0.009 & 1.5 $\pm$ 0.9 & 109 $\pm$ 17 \\
        & H$\alpha$   & 5.80 $\pm$ 1.06 & 3176 $\pm$ 98   & 6510.82 $\pm$ 7.55  & 0.001 $\pm$ 0.002 &-0.001 $\pm$ 0.002 & 0.1 $\pm$ 0.2 & 176 $\pm$ 49\\
        & H$\alpha$   & 2.35 $\pm$ 0.53 & 1873 $\pm$ 171  & 6559.21 $\pm$ 15.02 & 0.011 $\pm$ 0.005 & 0.018 $\pm$ 0.004 & 2.1 $\pm$ 0.5 & 29 $\pm$ 6\\
        & H$\alpha$   & 7.19 $\pm$ 0.51 & 1841 $\pm$ 124  & 6598.33 $\pm$ 3.84  & 0.019 $\pm$ 0.004 & 0.001 $\pm$ 0.003 & 1.9 $\pm$ 0.4 & 1 $\pm$ 5\\
        & H$\alpha$   & 6.65 $\pm$ 1.11 & 2778 $\pm$ 28   & 6625.73 $\pm$ 2.94  & 0.001 $\pm$ 0.003 & -0.001 $\pm$ 0.003 & 0.1 $\pm$ 0.3 & 169 $\pm$ 129\\
        &   \\
        \hline\\[-3pt]
        \multirow{17}{*}{\textbf{SDSS~0946}} 
        & continuum & - & - & -                                 & 0.004 $\pm$ 0.001  & -0.022 $\pm$ 0.002  & 2.3 $\pm$ 0.2 & 141 $\pm$ 2 \\
        & [O~I]     & 0.53 $\pm$ 0.02 & 269 $\pm$ 46  & 6300.30 & 0.041 $\pm$ 0.015 & 0.060 $\pm$ 0.018   & 7.2 $\pm$ 1.7 & 28 $\pm$ 6 \\
        & [S~III]   & 0.29 $\pm$ 0.26 & 274 $\pm$ 242  & 6312.06 & 0.077 $\pm$ 0.014 & 0.041 $\pm$ 0.020 & 8.7 $\pm$ 1.5 & 14 $\pm$ 6 \\
        & Si~II     & 0.15 $\pm$ 0.01 & 431 $\pm$ 57  & 6347.10 & -0.079 $\pm$ 0.021 & -0.005 $\pm$ 0.022 & 7.9 $\pm$ 2.1 & 92 $\pm$ 8 \\
        & [O~I]     & 0.20 $\pm$ 0.02 & 309 $\pm$ 44  & 6363.77 & 0.068 $\pm$ 0.019 & 0.046 $\pm$ 0.023  & 8.2 $\pm$ 2.0 & 17 $\pm$ 8 \\
        & Fe II     & 0.07 $\pm$ 0.02 & 213 $\pm$ 40  & 6369.46 & 0.078 $\pm$ 0.011 & 0.039 $\pm$ 0.019 & 8.7 $\pm$ 1.2 & 13 $\pm$ 5 \\
        & [Fe~X]    & 0.08 $\pm$ 0.01 & 444 $\pm$ 53  & 6374.51 & 0.065 $\pm$ 0.022 & 0.001 $\pm$ 0.024 & 6.5 $\pm$ 2.2 & 1 $\pm$ 10 \\
        & Fe II     & 2.08 $\pm$ 0.15 & 1874 $\pm$ 278  & 6516.08 & 0.009 $\pm$ 0.002 & -0.002 $\pm$ 0.004 & 1 $\pm$ 1 & 173 $\pm$ 14 \\
        & [N~II]    & 0.04 $\pm$ 0.03 & 151 $\pm$ 139  & 6548.05 & 0.007 $\pm$ 0.003 & 0.016 $\pm$ 0.004 & 1.8 $\pm$ 0.4 & 34 $\pm$ 6 \\
        & H$\alpha$ & 2.76 $\pm$ 0.57 & 549 $\pm$ 125  & 6562.81 & 0.012 $\pm$ 0.003 & 0.038 $\pm$ 0.005 & 3.9 $\pm$ 0.5 & 36 $\pm$ 2 \\
        & [N~II]    & 6.18 $\pm$ 0.54 & 979 $\pm$ 33  & 6583.46 & 0.004 $\pm$ 0.003 & 0.036 $\pm$ 0.002 & 3.4 $\pm$ 0.2 & 41 $\pm$ 3 \\
        & [S II]    & 0.60 $\pm$ 0.06 & 319 $\pm$ 36  & 6716.44 & -0.004 $\pm$ 0.024 & 0.005 $\pm$ 0.013 & 0.6 $\pm$ 1.8 & 64 $\pm$ 91 \\
        & [S II]    & 0.66 $\pm$ 0.06 & 302 $\pm$ 170  & 6730.81 & -0.017 $\pm$ 0.008 & 0.014 $\pm$ 0.006 & 2.2 $\pm$ 0.7 & 70 $\pm$ 8 \\
        & H$\alpha$ & 6.95 $\pm$ 0.66 & 2327 $\pm$ 120 & 6560.94 $\pm$ 1.48  & 0.005 $\pm$ 0.002 & 0.017 $\pm$ 0.002 & 1.7 $\pm$ 0.2 & 36 $\pm$ 3\\
        & H$\alpha$ & 13.93 $\pm$ 0.32 & 1418 $\pm$ 6  & 6603.37 $\pm$ 0.60  & -0.001 $\pm$ 0.002 & 0.016 $\pm$ 0.003 & 1.6 $\pm$ 0.3 & 46 $\pm$ 3\\
        &   \\
        \hline\\[-3pt]
        \multirow{14}{*}{\textbf{SDSS~1119}}
        & continuum & - & - & -                                    & -0.007 $\pm$ 0.001  & -0.006 $\pm$ 0.001  & 0.9 $\pm$ 0.1 & 109 $\pm$ 0.6  \\
        & [O~I]     & 0.251 $\pm$ 0.02 &  1010 $\pm$ 53  & 6300.30 & 0.029 $\pm$ 0.007 & -0.001 $\pm$ 0.001   & 2.9 $\pm$ 0.7 & 180 $\pm$1 \\
        & Si~II     & 0.08 $\pm$ 0.04 & 801 $\pm$ 322  & 6347.10 & 0.009 $\pm$ 0.007 & -0.005 $\pm$ 0.001 & 1,0 $\pm$ 0.6 & 166 $\pm$9 \\
        & [O~I]     & 0.09 $\pm$ 0.03 & 677 $\pm$ 209  & 6363.77 & 0.037 $\pm$ 0.024 & -0.019 $\pm$ 0.004  & 4.1 $\pm$ 2.1& 1666 $\pm$ 8 \\
        & Fe II     & 0.03 $\pm$ 0.04 &  136 $\pm$ 97  & 6369.46 & 0.074 $\pm$ 0.032 & 0.038 $\pm$ 0.029 & 8.3 $\pm$ 3 & 14 $\pm$ 10 \\
        & [Fe~X]    & 0.08 $\pm$ 0.02 & 785 $\pm$ 242  & 6374.51 & 0.228 $\pm$ 0.102 & -0.020 $\pm$ 0.001 & 23 $\pm$ 10 & 1777 $\pm$ 1\\
        & Fe II     & 0.12 $\pm$ 0.04 & 666 $\pm$ 388  & 6516.08 & 0.005 $\pm$ 0.003 & 0.018 $\pm$ 0.003 & 1.8 $\pm$ 0.3 & 36 $\pm$ 4\\
        & [N~II]    & 1.24 $\pm$ 0.37 & 435 $\pm$ 105  & 6548.05 & 0.005 $\pm$ 0.003 & -0.010 $\pm$ 0.007 & 1.1 $\pm$ 0.6 & 148 $\pm$ 11 \\
        & H$\alpha$ & 2.95 $\pm$ 0.32 & 811 $\pm$  85 & 6562.81 & -0.001 $\pm$ 0.003 & -0.005 $\pm$ 0.004 & 0.5 $\pm$ 0.4 & 133 $\pm$ 19 \\
        & [N~II]    & 2.10 $\pm$ 0.09 &  999 $\pm$ 7  & 6583.46 & 0.014 $\pm$ 0.002 & 0.008 $\pm$ 0.002 & 1.6 $\pm$ 0.2 & 15 $\pm$ 4 \\
        & H$\alpha$ & 2.56 $\pm$ 0.73 & 1086 $\pm$ 227 & 6538.89 $\pm$ 1.27  & 0.027 $\pm$ 0.002 & -0.002 $\pm$ 0.002 & 2.7 $\pm$ 0.2 & 178 $\pm$ 2\\
        & H$\alpha$ & 1.52 $\pm$ 0.70 & 3148 $\pm$ 187  & 6521.31 $\pm$ 14.25  & -0.003 $\pm$ 0.002 & -0.010 $\pm$ 0.003 & 1.0 $\pm$ 0.3 & 128 $\pm$ 5\\
        & H$\alpha$ & 2.54 $\pm$ 0.11 & 2771 $\pm$ 47 & 6610.04 $\pm$ 0.25  & -0.006 $\pm$ 0.002 & 0.015 $\pm$ 0.001 & 1.6 $\pm$ 0.1 & 56 $\pm 3$\\
        &   \\
        \hline\\[-3pt]
        \multirow{13}{*}{\textbf{SDSS~1536}}
        & continuum & - & - & -                                    & 0.003 $\pm$ 0.002  &  0.014 $\pm$ 0.001 & 1.4 $\pm$ 0.1 & 38 $\pm$ 3   \\
        & [O~I]      & 0.18 $\pm$ 0.02 & 1069 $\pm$ 154  & 6300.30 & -0.046 $\pm$ 0.010 & -0.004 $\pm$ 0.001  & 4.6 $\pm$ 1.0 & 92 $\pm$ 0.7  \\
        & [S~III]    & 0.02 $\pm$ 0.01 & 926 $\pm$ 257   & 6312.06 & -0.063 $\pm$ 0.012 & -0.005 $\pm$ 0.002 & 6.4 $\pm$ 1.2 & 92 $\pm$ 1\\
        & Si~II      & 0.02 $\pm$ 0.01 & 985 $\pm$ 274   & 6347.10 & -0.210 $\pm$ 0.038 & -0.006 $\pm$ 0.002 & 21.0 $\pm$ 3.8 & 91 $\pm$ 1\\
        & Fe II      & 1.56 $\pm$ 0.11 & 1177 $\pm$ 4    & 6516.08 & 0.007 $\pm$ 0.003 & -0.007 $\pm$ 0.003  & 1.0 $\pm$ 0.3 & 158 $\pm$ 8 \\
        & [N~II]     & 0.99 $\pm$ 0.09 & 698 $\pm$ 140   & 6548.05 & 0.004 $\pm$ 0.006 & -0.001 $\pm$ 0.003 & 0.4 $\pm$ 0.6 & 174 $\pm$ 20\\
        & H$\alpha$  & 3.19 $\pm$ 0.06 & 744 $\pm$ 38    & 6562.82 & -0.023 $\pm$ 0.003 & -0.003 $\pm$ 0.002 & 2.3 $\pm$ 0.3 & 94 $\pm$ 3\\
        & [N~II]     & 2.14 $\pm$ 0.10 & 975 $\pm$ 59    & 6583.46 & -0.001 $\pm$ 0.002 & -0.004 $\pm$ 0.003 & 0.4 $\pm$ 0.3 & 134 $\pm$ 13\\
        & H$\alpha$   & 3.60 $\pm$ 0.25 & 1261 $\pm$ 79  & 6487.10 $\pm$ 0.19  & -0.012 $\pm$ 0.002 & -0.005 $\pm$ 0.002 & 1.3 $\pm$ 0.2 & 101 $\pm$ 4\\
        & H$\alpha$   & 1.24 $\pm$ 0.24 & 4872 $\pm$ 118 & 6475.29 $\pm$ 7.74  & 0.004 $\pm$ 0.004 & 0.002 $\pm$ 0.002 & 0.4 $\pm$ 0.4 & 10 $\pm$ 13\\
        & H$\alpha$   & 1.72 $\pm$ 0.15 & 2123 $\pm$ 282 & 6617.97 $\pm$ 1.37  & -0.029 $\pm$ 0.003 & -0.009 $\pm$ 0.002 & 3.0 $\pm$ 0.3 & 99 $\pm$ 2\\
        & H$\alpha$   & 1.85 $\pm$ 0.10 & 1926 $\pm$ 278 & 6657.70 $\pm$ 1.33  & 0.001 $\pm$ 0.002 & -0.031 $\pm$ 0.002 & 3.1 $\pm$ 0.2 & 136 $\pm$ 2 \\[+3pt]
        \hline
        \multicolumn{9}{c}{}  
    \label{Tab:table_param_model}%
\end{longtable}

}


\end{appendix}

\end{document}